\documentclass[prd,aps,twocolumn,showpacs,superscriptaddress,nofootinbib]{revtex4-1}

\usepackage{bm}
\usepackage[colorinlistoftodos]{todonotes}
\usepackage{amsmath,amssymb}

\usepackage{graphicx,bm}
\usepackage[colorlinks,citecolor=blue]{hyperref}
\usepackage{breakurl}

\begin{document}

\title{Charged-Current Muonic Reactions in Core-Collapse Supernovae}

\author{Gang Guo}
\email{gangg23@gmail.com}
\affiliation{GSI Helmholtzzentrum f\"ur Schwerionenforschung,
  Planckstra{\ss}e~1, 64291 Darmstadt, Germany} 
\affiliation{Institute of Physics, Academia Sinica, Taipei, 11529, Taiwan}

\author{Gabriel Mart\'inez-Pinedo}
\affiliation{GSI Helmholtzzentrum f\"ur Schwerionenforschung,
  Planckstra{\ss}e~1, 64291 Darmstadt, Germany}
\affiliation{Institut
  f{\"u}r Kernphysik (Theoriezentrum), Technische Universit{\"a}t
  Darmstadt, Schlossgartenstra{\ss}e 2, 64298 Darmstadt, Germany}

\author{A. Lohs}
\affiliation{GSI Helmholtzzentrum f\"ur Schwerionenforschung,
  Planckstra{\ss}e~1, 64291 Darmstadt, Germany}
  
\author{Tobias Fischer}
\affiliation{Institute for Theoretical Physics, University of Wroc{\l}aw, 50-204 Wroc{\l}aw, Poland} 

\begin{abstract}
  The steady advance in core-collapse supernova simulations requires a
  more precise description of neutrino processes in hot and dense
  matter.  In this work, we study the rates of charged-current (CC)
  weak processes with (anti)muons in supernova matter. At the
  relativistic mean field level, we derive results for the rates of CC
  neutrino-nucleon reactions, taking into account full kinematics,
  weak magnetism and pseudoscalar terms, and $q^2$-dependent nucleon
  form factors in the hadronic current. In addition to muonic
  semileptonic processes we also consider purely leptonic
  processes. In particular, we show that inverse muon decay can
  dominate the opacities for low energy $\nu_\mu$ and $\bar\nu_e$ at
  densities $\gtrsim 10^{13}~\rm{g~ cm^{-3}}$.
\end{abstract}


\date{\today}
\maketitle

\section{Introduction}

The core-collapse of a massive star leads to the formation of a
neutron star and the subsequent supernova explosion. Most of the
gravitational binding energy of the neutron star is released in the form of neutrinos that in the neutrino-driven mechanism are
responsible for the ejection of the stellar
mantle~\cite{Janka:2012}. Consequently, neutrino processes in hot
and dense nuclear medium play crucial roles in many aspects of
core-collapse supernovae (CCSNe), in particular for the explosion
mechanism and the nucleosynthesis of heavy elements
\citep{Burrows.Reddy.Thompson:2006,Janka.Langanke.ea:2007,Burrows:2013,Martinez-Pinedo.Fischer.ea:2016}. The
successful explosion of core-collapse supernovae by the
neutrino-driven mechanism in three-dimensional simulations has
demonstrated the high relevance of neutrino physics as well as the
necessity of an accurate description of neutrino transport in hot and
dense protoneutron star (PNS) \citep{Melson.Janka.ea:2015}.

Neutrino processes in hot and dense matter have been well studied in
the literature
\cite{Schinder.Shapiro:1982,Bruenn:1985,Mezzacappa.Bruenn:1993,Hannestad.Raffelt:1998,Reddy.Prakash.Lattimer:1998,Reddy:1999hb,Horowitz:2002,Buras:2003wt,Horowitz.Schwenk:2006,Juodagalvis:2010pt,Roberts.Reddy.Shen:2012,Bartl:2014hoa,Horowitz.Caballero.ea:2017,Roberts.Reddy:2017,Bedaque:2018wns,Guo:2019cvs},
and their impacts on CCSN simulations have also been extensively
explored (see, e.g.,
Refs. \cite{Rampp:2002bq,Liebendoerfer:2005es,Buras:2006rp,Fischer_2010,MartinezPinedo:2012rb,OConnor:2015sgn,Bartl.Bollig.ea:2016,Fischer:2016boc,Roberts:2016lzn,Kotake:2018ypf,Fischer:2020kdt}).
However, neutrino CC processes considered for most of the studies are
limited to the lightest charged lepton, i.e., $e^\pm$. Due to a larger
rest mass, the production of $\mu^\pm$ was thought to be highly
suppressed and their role in SN dynamics was traditionally
ignored. Recently, the relevance of muons has been demonstrated in 2D
SN simulations \citep{Bollig:2017}. It showed that the formation of
muons in SN matter softens the equation of state (EOS), leads to
higher neutrino luminosities and mean energies, and therefore
facilitates neutrino-driven explosions. It should be pointed out that the
production of $\mu^\pm$, especially the accumulation of net
$\mu$-lepton number, is closely related to the CC reactions of
$\nu_\mu$($\bar\nu_\mu$), which are created via thermal pair processes
like $e^- + e^- \to \nu_\mu+\bar\nu_\mu$ and
$N+N \to N+N+\nu_\mu+\bar\nu_\mu$, and affected by neutrino transport in SN
matter.

To be specific, the more abundant electrons, compared to positrons,
can lead to an excess of $\mu^-$ than $\mu^+$ via leptonic weak
processes like $\nu_\mu/\bar\nu_e+e^-\to
\nu_e/\bar\nu_\mu+\mu^-$. This is also aided by the semileptonic
process, $\nu_\mu+n\to p+\mu^-$, which is more favoured than
$\bar\nu_\mu+p\to n+\mu^+$ since neutrons are in higher energy states
than protons. Due to muon number conservation, the initial excess of
$\mu^-$ is compensated by an excess of $\bar\nu_\mu$. With a more
abundant flux and a lower neutral-current (NC) scattering cross
section with nucleon, more $\bar\nu_\mu$ diffuse out of the
protoneutron star, leading to a gradual buildup of net muon number,
i.e., muonization. The appearance of muonization not only affects the
neutrino spectra of all flavors and enhances the explosion ability,
but may also play a non-negligible role in the subsequent neutron
star/black hole formation~\citep{Bollig:2017}. 
                             
In this work, we aim to study the reaction rates of all the relevant
weak processes involving $\mu^-$ or $\nu_\mu$ (see
Table~\ref{tab:mu_reaction}), which are required as input in numerical
simulations for a consistent description of muonization and neutrino
transport. It has been found that weak magnetism can enhance/suppress
the neutrino/antineutrino opacities significantly and thus affects the
neutrino spectra and SN dynamics~\citep{Horowitz:2002}. A relativistic
treatment of both full kinematics (nuclear recoil) and weak magnetism
for CC $\nu_e$($\bar\nu_e$)-nucleon reactions have been
studied~\cite{Roberts.Reddy:2017,Fischer:2020kdt}, with its impact
recently explored in symmetric CCSN
simulation~\cite{Fischer:2020kdt}. For semileptonic reactions
involving $\mu^\pm$ with large energy-momenta transfer, pseudoscalar
coupling term in the hadronic weak current, which is normally
neglected for $\nu_e$ reactions, is found as important as weak
magnetism. Besides, the effects of nucleon form factors become
comparable as energy-momenta transfer increases and need to be
considered on the same footing as weak magnetism and pseudoscalar
corrections. Hence, we extend the formalisms presented
in~\cite{Fischer:2020kdt} to include weak magnetism, pseudoscalar term
and $q^2$-dependent form factors in the hadronic current for CC
$\nu_\mu$-nucleon reaction. Treating nucleons at the mean field level
for the semileptonic processes, we obtain the opacities with full
kinematics in a fully relativistic formalism. Correlations among
nucleons beyond the mean field level have been investigated in
Refs. \cite{Burrows:1998cg,Reddy.Prakash.Lattimer:1998,Burrows:1999ek,Reddy:1999hb,Horowitz:2003yx,Roberts.Reddy.Shen:2012}
and are neglected in this work.  Purely leptonic processes have
already been studied in the
literature~\citep{Schinder.Shapiro:1982,Mezzacappa.Bruenn:1993}. We
will compare them with the semileptonic processes and discuss their
relevances to $\mu^\pm$ production and neutrino transport. As
  muonic processes also contribute to the opacity of $\bar{\nu}_e$
  neutrinos we also update the description of inverse neutron decay
  that has recently been studied in Ref.~\cite{Fischer:2020kdt}. We also note that the relevance of pions in hot dense nuclear matter has recently been explored and weak reactions involving pions and muons can be another important opacity source for $\nu_\mu$ and $\bar\nu_\mu$ \cite{Fore:2019wib}. Here we only focus on the standard muonic reactions listed in Table~\ref{tab:mu_reaction} and briefly mention the possible role of pionic reactions.     
  
The paper is organised as follows. We present the formalisms to
calculate the rates of purely leptonic processes in Sec.~\ref{sec:lep}
and the rates of the semileptonic processes in
Sec.~\ref{sec:semilep}. In Sec.~\ref{sec:results}, we discuss the
effects of weak magnetism, pseudoscalar term and form factors in
semileptonic processes involving $\mu^\pm$, and then compare the
contributions from different leptonic and semileptonic processes. We
come to the conclusion in Sec.~\ref{sec:summary}.

\begin{table}[htbp] 
\centering 
\caption{Weak reactions with $\mu^\pm$ or $\nu_\mu$ considered in this work. We also consider inverse neutron decay as opacity source for $\bar\nu_e$. \label{tab:mu_reaction}}
\begin{ruledtabular}
\begin{tabular}{ll}    
Leptonic reactions & Semileptonic reactions \\    
\hline
 $\nu_\mu + e^- \to \nu_\mu + e^-$ (a) & $\nu_\mu + n \to \mu^- + p$ (f)\\ 
 $\bar\nu_\mu + e^- \to \bar\nu_\mu+ e^-$ (b)& $\bar\nu_\mu + p \to
                                               \mu^+ + n$ (g) \\
 $\nu_\mu + e^- \to \nu_e + \mu^-$ (c) &  $\bar\nu_e + p +e^- \to n$~~(h) \\
 $\bar\nu_e + e^- \to \nu_\mu + \mu^-$ (d)&   \\ 
 $\nu_\mu + e^- + \bar\nu_e \to \mu^-$ (e) &  \\
\end{tabular}
\end{ruledtabular}
\end{table}   

\section{Leptonic reactions}  
\label{sec:lep} 

\subsection{scattering}

For neutrinos scattering with charged leptons, $\nu_1 + l_2^- \to \nu_3 + l_4^-$, the spin-averaged matrix elements can be expressed in a general form as 
\begin{align}
\langle |{\cal M}|^2 \rangle =& \lambda_1 (p_1 \cdot p_2) (p_3 \cdot p_4) + \lambda_2 (p_1\cdot p_4)(p_2 \cdot p_3) \nonumber \\ 
&+\lambda_3(p_1 \cdot p_3),  
\label{eq:msq_lep}
\end{align}
where $p_{1,2,3,4}$ are the four-momenta of $\nu_1$ (particle `1'),
$l_2$ (particle `2'), $\nu_3$ (particle `3') and $l_4$ (particle `4'),
respectively. The coefficients $\lambda_i$ are shown in
Appendix~\ref{sec:append_lep} for different processes in
Table~\ref{tab:mu_reaction}.
The scattering kernel for this general reaction can be
expressed as~\cite{Bruenn:1985}
\begin{align}
R^{\rm in}(E_1, E_3,\mu) =& 2 \int \frac{d^3p_2}{(2\pi)^3}\frac{d^3p_4}{(2\pi)^3} \frac{\langle |{\cal M}|^2 \rangle}{16E_1E_2E_3E_4} \nonumber \\
& \times (2\pi)^4 \delta^{(4)}(p_1+p_2-p_3-p_4) \nonumber \\
& \times f_2(E_2)[1-f_4(E_4)]  \nonumber \\
=& \lambda_1 R_1 +\lambda_2 R_2 + \lambda_3 R_3,  
\label{eq:kernel_lep}
\end{align}
where $E_i$ are the relativistic energies, $f_{2, 4}$ are the Fermi
distribution functions of the charged leptons, $\mu=\cos\theta_{13}$
with $\theta_{13}$ the angle between $\bm{p}_{1, 3}$, and $R_{1,2,3}$
are contributions from different terms of
$\langle |{\cal M}|^2 \rangle$ shown in
Eq.~(\ref{eq:msq_lep}). Due to rotational invariance, the
  scattering kernel only depends on the relative angle between the
  incoming and outgoing neutrinos. It can also be checked that both
  the matrix element and the phase space integral do not depend on the
  azimuthal angle.

The $R_i$ can be solved analytically up to a remaining integral over $E_2$ as \cite{Yueh:1976A,Yueh:1976B,Schinder.Shapiro:1982,Mezzacappa.Bruenn:1993,Lohs:2015} 
\begin{align}
R_i =& \frac{1}{16\pi \Delta^5} \int_{E_-}^{\infty} dE_2 f_2(E_2)[1-f_4(E_4)] (A_{i}E_2^2 + B_{i}E_2+C_i) \nonumber \\
=& \frac{1}{16\pi \Delta^5} (A_{i} I_2 + B_{i} I_1 + C_{i} I_0), 
\label{eq:R_i}
\end{align}
where $E_-$ and $\Delta$ are given by 
\begin{equation}
\begin{aligned}
E_- =& \frac{1}{2}\Bigg\{(E_3-E_1)(1+k) \\ 
&+\Delta \sqrt{\bigg[(1+k)^2+\frac{2m_2^2}{E_1E_3(1-\mu)}\bigg]} \Bigg\}, \\
\Delta =& \sqrt{E_1^2-2E_1E_3\mu+E_3^2}, 
\end{aligned}  
\end{equation}
with 
\begin{align}
k= \frac{Q}{E_1E_3(1-\mu)}, \;\; Q= \frac{1}{2}(m_4^2-m_2^2).
\label{eq:kQ} 
\end{align}
Similarly, the coefficients $A_i$, $B_i$, and $C_i$ are all functions
of $E_{1, 3}, \mu$ and $m_{2, 4}$, and are presented in
Appendix~\ref{sec:append_lep}. The functions $I_{s=0,1,2}$ are defined
as
\begin{align}
I_s = \int_{E_-}^\infty dE_2 E_2^s f_2(E_2)[1-f_4(E_1+E_2-E_3)],  
\label{eq:I_s}
\end{align}
and they can be expressed in terms of the Fermi-Dirac integrals which are more convenient to compute numerically,
\begin{subequations}
\label{eq:hatI}
\begin{equation}
\begin{aligned}
I_0 =&~ T\widehat I_0(\eta',\eta,y)\\
=&~ T f_\gamma(\eta'-\eta) [F_0(\eta'-y)-F_0(\eta-y)], 
\end{aligned}
\end{equation}
\begin{equation}
\begin{aligned}
I_1 =&~ T^2 \widehat I_1(\eta',\eta,y) \\ 
=&~ T^2 f_\gamma(\eta'-\eta) \big\{ [F_1(\eta'-y)-F_1(\eta-y)]  \\ 
  &+ y[F_0(\eta'-y)-F_0(\eta-y)] \big\}, 
\end{aligned}
\end{equation}
\begin{equation}
\begin{aligned}
I_2 =&~T^3\widehat I_2(\eta',\eta,y)\\
=&~T^3 f_\gamma(\eta'-\eta) \big\{ [F_2(\eta'-y)-F_2(\eta-y)]  \\ 
&+ 2y[F_1(\eta'-y)-F_1(\eta-y)]  \\
& + y^2[F_0(\eta'-y)-F_0(\eta-y)] \big\}, 
\end{aligned} 
\end{equation}
\end{subequations} 
where the Fermi-Dirac integrals $F_n(z)$ and the function $f_\gamma(z)$ are defined by 
\begin{align}
F_n(z) & = \int_0^\infty dx \frac{x^n}{\exp(x-z)+1}, \\
f_\gamma(z) &= \frac{1}{\exp(z)-1}, \label{eq:fg}
\end{align}
and the coefficients $y$, $\eta$, and $\eta'$ are given by  
\begin{align}
y = \frac{E_-}{T}, \;\; \eta=\frac{\mu_2}{T}, \;\; \eta'=\frac{E_3-E_1+\mu_4}{T},  
\end{align}
with $T$ the temperature and $\mu_{2,4}$ the relativistic chemical
potentials including the rest mass.
Note that $F_0(z)$ has an exact expression $\ln(1+e^z)$. For
  $n\ge 1$, there are approximate expressions available for $F_n(z)$
  which are valid for $z\gg0$ and $z\ll0$ and can reproduce the exact
  results within 20\% at $z\approx 0$
  \cite{Fuller:1085zz,Martinez-Pinedo:2014koa}. In this work we
  calculate $F_n$ ($n\ge 1$) numerically. We also note that
  $f_\gamma(\eta'-\eta)$ in Eq.~\eqref{eq:fg} is divergent for elastic
  scattering with $\eta=\eta'$. For this special case, we have
\begin{subequations}
\begin{equation}
\begin{aligned}
I_0 =& T F'_0(z)\big|_{z=\eta-y}\;,  
\end{aligned}
\end{equation}
\begin{equation}
\begin{aligned}
I_1 =& T^2 \big[F'_1(z)+yF'_0(z)\big]\big|_{z=\eta-y}\;,
\end{aligned}
\end{equation}
\begin{equation}
\begin{aligned}
I_2 =& T^3 \big[F'_2(z) + 2y F'_1(z) + y^2F'_0(z)\big]_{z=\eta-y}\;, 
\end{aligned} 
\end{equation}
\end{subequations}
where $F'_n(z)\equiv dF_n(z)/dz$. For $n\ge1$, $F'_n(z)=nF_{n-1}(z)$.

Once the scattering kernels are known, neutrino opacity or inverse
mean free path due to scattering with leptons can be obtained by
integrating over the phase space of the final-state neutrino as
\begin{align}
\chi(E_1)= \int \frac{d^3p_3}{(2\pi)^3} R(E_1, E_3,\mu)[1-f_3(E_3)].   
\end{align}
The scattering kernel for the inverse process, $\nu_1 + l^-_2 \leftarrow
\nu_3 + l^-_4$, can be obtained via detailed balance:

\begin{equation}
  \label{eq:inv}
  R^{\rm{out}}(E_1, E_3,\mu) = \exp\left\{\frac{E_3-E_1-\Delta\mu_{24}}{T}\right\} R^{\rm in}(E_1, E_3,\mu)
\end{equation}
with $\Delta\mu_{24}=\mu_2-\mu_4$.

\subsection{inverse muon decay}

For inverse decay, we consider a process like
$\nu_1 + l_2 + \nu_3 \to l_4$ in which two neutrinos are
absorbed. Similarly to scattering, the matrix elements can be
expressed as
\begin{align}
\langle |{\cal M}|^2 \rangle =& \lambda_{A1} (p_1 \cdot p_2) (p_3 \cdot p_4) + \lambda_{A2} (p_1\cdot p_4)(p_2 \cdot p_3) \nonumber \\ 
&+\lambda_{A3}(p_1 \cdot p_3),  
\label{eq:msq_lep_inv}
\end{align} 
where the coefficients $\lambda_{A1,A2,A3}$ are given in Appendix~\ref{sec:append_lep}.

The kernel for inverse decay can be written as
$R_A(E_1, E_3,\mu)=\lambda_{A1} R_{A1}+\lambda_{A2}
R_{A2}+\lambda_{A3} R_{A3}$, and $R_{Ai}$ are given by
\begin{align}
R_{Ai} =& \frac{1}{16\pi\Delta_A^5} \int_{E_{A-}}^{E_{A+}} dE_2 f_2(E_2)[1-f_4(E_4)]  \\
  & \times (A_{Ai}E_2^2 + B_{Ai}E_2+C_{Ai}) \Theta(k-k_0) \nonumber \\
=& \frac{1}{16\pi \Delta_A^5} (A_{Ai} I_{A2} + B_{Ai} I_{A1} + C_{Ai} I_{A0}) \Theta(k-k_0) , 
\label{eq:R_Ai}
\end{align}  
where 
\begin{equation}
\begin{aligned}
E_{A\pm} =& \frac{1}{2}\Bigg\{(E_3+E_1)(k-1)   \\ 
&\pm \Delta_A \sqrt{\bigg[(1-k)^2-\frac{2m_2^2}{E_1E_3(1-\mu)}\bigg]} \Bigg\},   \\
\Delta_A =& \sqrt{E_1^2+2E_1E_3\mu+E_3^2}, \\
k_0 =& \frac{m_2+m_4}{m_4-m_2}, 
\end{aligned}
\end{equation}
with $k$ and $Q$ given in Eq.~(\ref{eq:kQ}). The coefficients $A_{Ai}$, $B_{Ai}$, and $C_{Ai}$ are presented in Appendix.~\ref{sec:append_lep}, and 
\begin{subequations}
\begin{equation}
\begin{aligned}
I_{A0} =&~ T[\widehat I_0(\eta'_A,\eta,y_-)-\widehat I_0(\eta'_A,\eta,y_+)],
\end{aligned}    
\end{equation}
\begin{equation}
\begin{aligned}
I_{A1} =&~ T^2[\widehat I_1(\eta'_A,\eta,y_-)-\widehat I_1(\eta'_A,\eta,y_+)],
\end{aligned}    
\end{equation}
\begin{equation}
\begin{aligned}
I_{A2} =&~ T^3[\widehat I_2(\eta'_A,\eta,y_-)-\widehat I_2(\eta'_A,\eta,y_+)],
\end{aligned}    
\end{equation}
\end{subequations}
with $\widehat I_{0,1,2}$ introduced in Eq.~\eqref{eq:hatI} and $y_\pm$, $\eta$, and $\eta'_A$ given by  
\begin{align}
y_\pm = \frac{E_{A\pm}}{T}, \;\; \eta=\frac{\mu_2}{T}, \;\; \eta'_A=\frac{-E_3-E_1+\mu_4}{T}.  
\end{align}

Opacity for $\nu_{1}$ due to inverse decay can be expressed as
\begin{equation} 
\chi(E_1) = \int \frac{d^3p_3}{(2\pi)^3} R_A(\mu, E_1, E_3)f_3(E_3).   
\end{equation}
Similarly, one can obtain the opacity for $\nu_3$ by replacing
$d^3p_3$ with $d^3p_1$ and $f_1(E_1)$ with $f_3(E_3)$ in the above
equation.

The kernel for the decay process,
$l_4 \to \nu_1 + l_2 + \nu_3$, can be obtained from detailed balance:
\begin{equation}
  \label{eq:2}
  R_D(E_1,E_3,\mu)= \exp\left\{-\frac{E_1+E_3+\Delta \mu_{24}}{T}\right\} R_A(E_1, E_3,\mu).
\end{equation}

\section{Semileptonic reactions}  
\label{sec:semilep}
 
\subsection{neutrino absorption}
\label{sec:v_ab} 

We consider semileptonic CC reactions of the form
\begin{equation}
 \label{eq:semi_reaction}
 \nu_1+N_2\rightarrow l_3+N_4,
\end{equation}
where $N_{2,4}$ stand for the initial and final state nucleons.

We take relativistic dispersion relations for nucleons in the nuclear
medium, which can be parametrized in terms of effective masses $m^*$
and mean field potentials $U$ as
\begin{equation}
 \label{eq:2.3}
 E_{2,4}=\sqrt{m_{2,4}^{*2}+|\bm{p}_{2,4}|^2}+U_{2,4},
\end{equation}
with the four-momentum of nucleons
$p_{2,4}=(E_{2, 4}, \bm{p}_{2, 4})$.  The interaction potentials and
effective masses can be deduced from relativistic mean field theory,
based on which equation of state (EoS) of hot and dense nuclear matter
can be derived
\cite{Shen:1998gq,Typel:1999yq,Typel:2005ba,Hempel:2010mc,Shen:2010pu,Typel:2010sy,Furusawa:2011wh,Shen:2011kr,Shen:2011fc,Hempel:2012mk,Steiner:2013rk}.
The nuclear EoS can also be studied based on nonrelativistic
parameterizations of the nuclear potentials, such as the widely used
Lattimer and Swesty (LS) EoS~\cite{Lattimer.Swesty:1991}. In this
case, nucleons take nonrelativistic energy-momentum relation as
$E_{\rm{NR}} = |\bm{p}|^2/(2m^*) + U$. When transferred to
relativistic form, $E \simeq \sqrt{m^{*2} + |\bm{p}|^2}+m-m^*+U$. Note
that within the LS EoS, the Landau effective masses of nucleons
are simply the bare masses.

For the energies considered, the neutrino-nucleon interaction is
described by a current-current interaction
\begin{equation}
 \label{eq:lagrangian_semi}
 \mathcal{L} = \frac{G}{\sqrt{2}} l_\mu j^\mu, 
\end{equation}
where $G=G_{\text{F}} V_{ud}$ for CC processes and $G=G_{\rm F}$ for
NC processes, with $G_{\rm F}$ the Fermi coupling constant and
$V_{ud}$ the up-down entry of the Cabibbo-Kobayashi-Maskawa matrix. The leptonic current is
\begin{equation}
  \label{eq:leptoncurrent}
  l_\mu = \bar l_3\gamma_\mu (1-\gamma_5)\nu_1,  
\end{equation}
and the effective hadronic current takes a general form, 
\begin{equation}
  \label{eq:hadroncurrent}
  \begin{aligned}
  j^\mu = \bar{\psi}_4 \Bigg\{ & \gamma^\mu \Big[G_V(q^2)-G_A(q^2)\gamma^5\Big] \\   
   & + \frac{i F_2(q^2)}{2 M_N} \sigma^{\mu\nu} q^*_\nu - \frac{G_P(q^2)}{M_N} \gamma^5 q^{*\mu}\Bigg\} \psi_2, 
  \end{aligned}
\end{equation}
where $\psi_{2,4}$ are the Dirac spinors of nucleons, $q = p_{1} -
p_{3}=p_4-p_2$ is the momentum transferred to the nucleon, and $M_N$
is the nucleon mass taken to be the average nucleon bare mass with
$M_N=(m_n+m_p)/2$. Note that for weak magnetism and pseudoscalar term,
we introduce $q^* = p_4^*-p_2^*$ with $p^*_{2,4} = (E^*_{2,4}, \bm{p}_{2,4})= \Bigl(\sqrt{|\bm{p}_{2, 4}|^2+m_{2,4}^{*2}}, \bm{p}_{2,4}\Bigr)$
as required by conservation of the weak
vector current \cite{Leinson_2001,Leinson:2002bw,Roberts.Reddy:2017,Fischer:2020kdt}. The vector term, axial vector term, weak magnetism as well as pseudoscalar term, are all characterized by $q^2$-dependent
coupling strengths or form factors as 
\begin{equation}
\label{eq:form-factor} 
\begin{aligned}
 G_V(q^2) &= \frac{  g_V\Big[1-\frac{q^2(\gamma_p-\gamma_n)}{4M_N^2}\Big] }{  \big(1-\frac{q^2}{4M_N^2}\big) \big(1-\frac{q^2}{M_V^2}\big)^2 }, \\
 G_A(q^2) &= \frac{ g_A }{ \big(1-\frac{q^2}{M_A^2}\big)^2 }, \\
 F_2(q^2) &=
\frac{\gamma_p - \gamma_n -1}{ \big(1-\frac{q^2}{4M_N^2}\big)\big(1-\frac{q^2}{M_V^2}\big)^2 }, \\
G_P(q^2) &= \frac{2 M_N^2 G_A(q^2)}{ m_\pi^2-q^2}, 
\end{aligned}    
\end{equation}
where $\gamma_{p, n}$ are the magnetic moments of protons and neutrons
with $\gamma_p \simeq 2.793$ and $\gamma_n \simeq -1.913$, the vector and
axial vector coupling constants are $g_V= 1$ and $g_A\simeq 1.27$, and
$M_{V}\simeq 840$ MeV, $M_{A}\simeq 1$ GeV, and
$m_\pi \simeq 139.57$~MeV are the vector mass, the axial mass, and the
pion mass, respectively.
As commonly taken in the literature, the
above form factors can be approximated by $g_{V,A}$, $\gamma_p -\gamma_n-1$ and
$g_P =2m_N^2 g_A/(m_\pi^2-q^2)$ if the momentum transfer dependence can be
neglected. Note that the pole in $g_P$ at $q^2 = m_\pi^2$ can not be encountered for muonic reactions since $q^2=(p_1-p_3)^2=m_{\mu}^2-2p_1\cdot p_3<m_{\mu}^2<m_\pi^2$.
The relevance of pseudoscalar term for $\nu_\mu$ reactions
can be understood since $q^2$ can get close to $m_\mu^2$,
resulting in a large value of $g_P(q^2)$. We will explore the effects
of these interaction terms and form factors later in this work.

Different from scattering process for which one needs to compute the
scattering kernel, for neutrino absorption we are interested in the
opacity or inverse mean free path:\footnote{We note that in Eq. (5) of
  our previous paper \cite{Fischer:2020kdt}, $E_{2,4}$ should be
  replaced by $E_{2,4}^*$. However, they were correctly treated in our
  numerical calculations and studies therein are unaffected.}
\begin{align}
 \chi(E_1)&=  2 \int \frac{d^3\bm{p}_2}{\left(2\pi\right)^3} \int \frac{d^3\bm{p}_3}{\left(2\pi\right)^3} \int \frac{d^3\bm{p}_4}{\left(2\pi\right)^3}
 \frac{\langle\left|\mathcal{M}\right|^2\rangle}{16E_1E_2^*E_3E_4^*} \nonumber\\
  \times & \left(2\pi\right)^4\delta^{(4)}\!\left(p_1+\!p_2-\!p_3-\!p_4\right) f_2 \left(1-f_3\right) \left(1-f_4\right),
\label{eq:Msq_semi}   
\end{align}
where the square of the amplitude can be written as 
\begin{equation}
\begin{aligned}
  \label{eq:matrixedecomp}
  \langle |\mathcal{M}|^2\rangle = & \langle |\mathcal{M}|^2\rangle_{VV} \pm \langle |\mathcal{M}|^2\rangle_{VA} + \langle|\mathcal{M}|^2\rangle_{AA} \\ 
  &+ \langle |\mathcal{M}|^2\rangle_{VF} \pm\langle |\mathcal{M}|^2\rangle_{AF} + \langle|\mathcal{M}|^2\rangle_{FF} \\
  &+ \langle |\mathcal{M}|^2\rangle_{AP} + \langle |\mathcal{M}|^2\rangle_{PP},
 \end{aligned}
\end{equation}
corresponding to contributions from the vector ($V$), axial vector
($A$), weak magnetism ($F$), pseudoscalar ($P$) terms, and their
interferences. They are functions of momenta of the initial and the
final particles, see Appendix~\ref{sec:append_semi}. The
``$+$''(``$-$'') sign applies to neutrino(antineutrino) absorption,
which is required by $CPT$ symmetry. As demonstrated in
Ref.~\cite{Horowitz:2002}, the $AF$ interference term,
$\pm\langle |\mathcal{M}|^2\rangle_{AF}$, increases (decreases) the
neutrino (antineutrino) opacities. The $AP$ interference term,
$\langle |\mathcal{M}|^2\rangle_{AP}$, contributes negatively, leading
to a suppressed opacity for both neutrinos and antineutrinos. Note
that there is no interference between $P$ and $V$/$F$ terms.

To obtain $\chi(E_\nu)$ for neutrino transport, we can either first
integrate out all the angles analytically in Eq.~(\ref{eq:Msq_semi}),
and then do a remaining 2D integral over energies numerically
\cite{Lohs:2015}, or directly perform a 4D numerical integration after
applying the energy-momentum $\delta$-function without any further
analytical integration. We will do both ways independently, and show
they give rise to similar results within $\sim 10^{-3}$ for all the
conditions considered.

\subsubsection{4D integrals} 
In the 4D integrals, we choose $E_2$, $|{\bm q}|=|{\bm p}_4-\bm{p}_2|$, $\cos\theta_{q2}$ and $\phi_{q2}$ as the integration variables, where $\theta_{q2}$ and $\phi_{q2}$ are the polar and azimuthal angles of $\bm{p}_2$ with respect to $\bm{q}$. The opacity due to absorption on nucleons can be expressed as 
\begin{align}
\chi(E_1) =& 2\int d\phi_{q2} \int\cos\theta_{q2} \int d|{\bm q}| \int dE_2 \langle |{\cal M}|^2 \rangle  \nonumber \\
& \times f_2(1-f_3)(1-f_4) \frac{|{\bm q}| |{\bm p_2}|}{16 (2\pi)^4 E_1^2 E_4^*}.  
\end{align}
The determination of the integration bounds is discussed in
Appendix~\ref{sec:bound_cap}. In this work, all the 4D integrals are
solved via a Monte Carlo algorithm encoded in the CUBA
library~\citep{Hahn:2005}.

\subsubsection{2D integrals} 
To perform the integration over angles analytically, one can
sort the total matrix elements into different four-momenta products:
\begin{widetext}
\begin{equation}
\label{eq:semi_amp_decomp}   
\begin{aligned}
 \left\langle |\mathcal{M}|^2\right\rangle 
 = & \mathcal{A} M_\mathcal{A} +\mathcal{B} M_\mathcal{B} + \dots + \mathcal{K}_{\rm{tot}} M_\mathcal{K} + \mathcal{L}_{\rm{tot}} M_\mathcal{L} 
 = (4G)^2 \left[\mathcal{A}\left(p_1\cdot p_2^*\right)\left(p_3\cdot p_4^*\right) \right. \\
 & + \mathcal{B}\left(p_1\cdot p_4^*\right)\left(p_3\cdot p_2^*\right) +\mathcal{C}\left(p_1\cdot p_2^*\right)^2\left(p_1\cdot p_3\right)
 +\mathcal{D}\left(p_1\cdot p_2^*\right)\left(p_1\cdot p_3\right)^2  
 +\mathcal{E}\left(p_1\cdot p_2^*\right)^2 \\
 &+\mathcal{F}_{\rm{tot}}\left(p_1\cdot p_3\right)^2 +\mathcal{H}\left(p_1\cdot p_2^*\right)\left(p_1\cdot p_3\right)
 +\left. \mathcal{J}_{\rm{tot}}\left(p_1\cdot p_2^*\right) +\mathcal{K}_{\rm{tot}}\left(p_1\cdot p_3\right) + \mathcal{L}_{\rm{tot}}\right],
\end{aligned}        
\end{equation}
with 
\begin{equation}
\begin{aligned}
& \mathcal{F}_{\rm{tot}} = \mathcal{F} + \frac{\mathcal{F}^{PP}}{(m_\pi^2-q^2)^2},\;\; 
\mathcal{J}_{\rm{tot}} = \mathcal{J} + \frac{\mathcal{J}^{AP}}{m_\pi^2-q^2}, \\ 
& \mathcal{K}_{\rm{tot}} = \mathcal{K} + \frac{\mathcal{K}^{AP}}{m_\pi^2-q^2} + \frac{\mathcal{K}^{PP}}{(m_\pi^2-q^2)^2}, \;\;
\mathcal{L}_{\rm{tot}} = \mathcal{L} + \frac{\mathcal{L}^{AP}}{m_\pi^2-q^2} + \frac{\mathcal{L}^{PP}}{(m_\pi^2-q^2)^2}.
\end{aligned} 
\end{equation}

\end{widetext}

Neglecting firstly the form factor dependences, the coefficients $\mathcal{A},\mathcal{B},\dots,\mathcal{K}_{\rm{tot}}$, and $\mathcal{L}_{\rm{tot}}$ are independent of momenta (see Appendix~\ref{sec:append_semi}), and can be taken out of the integral.
With the above notation, the opacity can be expressed as 
\begin{equation}
\begin{aligned}
 \label{eq:imfp-I}           
 \chi(E_1) =& \frac{G^2}{4\pi^3}\frac{1}{E_1^2}\int\limits_{E_{3-}}^{E_{3+}}dE_3 \int\limits_{E_{2-}}^{E_{2+}}dE_2 f_2(1-f_3)(1-f_4) \\  
 & \times \Big[\mathcal{A}I_\mathcal{A}+\mathcal{B}I_\mathcal{B}+\dots + \mathcal{E}I_\mathcal{E} + \mathcal{H} I_\mathcal{H} \\
 & +(\mathcal{F} I_\mathcal{F}+\mathcal{F}^{PP} I_\mathcal{F}^{PP}) + (\mathcal{J} I_\mathcal{J} + \mathcal{J}^{AP} I_J^{AP})  \\
 & +(\mathcal{K} I_\mathcal{K}+\mathcal{K}^{AP} I_\mathcal{K}^{AP} + \mathcal{K}^{PP} I_\mathcal{K}^{PP}) \\
 & +(\mathcal{L} I_\mathcal{L}+\mathcal{L}^{AP} I_\mathcal{L}^{AP} + \mathcal{L}^{PP} I_\mathcal{L}^{PP}) \Big],     
\end{aligned}
\end{equation}
where $I_\mathcal{X}$ ($\mathcal{X}=\mathcal{A}, \mathcal{B}, ...$) are integrals over the angles and are given by
\begin{equation}
\begin{aligned}
\label{eq:I_x} 
 I_\mathcal{X} &= \frac{\bar p_1\bar p_2\bar p_3\bar p_4}{4\pi^2} \!\!\int\!\! d\Omega_2 d\Omega_3 d\Omega_4 dE_4 M_\mathcal{X} \delta^{(4)}\!\left(p_1\!+p_2\!-p_3\!-p_4\right) \\
 &=\frac{\bar p_1\bar p_2\bar p_3\bar p_4}{4\pi^2} \!\!\int\!\! d\Omega_2 d\Omega_3 d\Omega_4 M_\mathcal{X} \delta^{(3)}\!\left(\bm{p}_1\!+\!\bm{p}_2\!-\!\bm{p}_3\!-\!\bm{p}_4\right),
\end{aligned}  
\end{equation} 
with $\bar{p}_i = |\bm{p}_i|$ and $M_\mathcal{X}$ introduced in Eq.~\eqref{eq:semi_amp_decomp}. Similarly, $I_\mathcal{X}^{AP,PP}$ are given by 
\begin{equation}
\begin{aligned}
I_\mathcal{X}^{AP} =\frac{\bar p_1\bar p_2\bar p_3\bar p_4}{4\pi^2} \!\!\int\!\! d\Omega_2 & d\Omega_3 d\Omega_4 \frac{M_\mathcal{X}}{m_\pi^2-q^2} \\
  & \times \delta^3\!\left(\bm{p}_1\!+\!\bm{p}_2\!-\!\bm{p}_3\!-\!\bm{p}_4\right), \\ 
I_\mathcal{X}^{PP} =\frac{\bar p_1\bar p_2\bar p_3\bar p_4}{4\pi^2} \!\!\int\!\! d\Omega_2 & d\Omega_3 d\Omega_4 \frac{M_\mathcal{X}}{(m_\pi^2-q^2)^2} \\ 
 & \times \delta^3\!\left(\bm{p}_1\!+\!\bm{p}_2\!-\!\bm{p}_3\!-\!\bm{p}_4\right). 
\end{aligned}
\end{equation}
All the $I_\mathcal{X}, I_\mathcal{X}^{AP,PP}$ can be computed analytically \cite{Lohs:2015}, and their expressions are presented in Appendix~\ref{sec:append_semi}. The bounds $E_{2\pm}$ and $E_{3\pm}$ 
can be obtained numerically from energy-momenta conservation, and only the kinematically allowed regions contribute to the opacities, see Appendix~\ref{sec:append_bounds}.  
                       
As mentioned above, for neutrinos with energies $E_1 \gtrsim 100$ MeV
or reactions involving $\nu_\mu(\bar\nu_\mu)$, the suppression of the
rates due to form factor will become significant. Considering the
$q^2$-dependent coupling constants, see Eq.~(\ref{eq:form-factor}),
the analytical integrations over angles become nontrivial. One
possible way is to expand the form factors in powers of
$q^2/M^2_{V,A,N}$, and do the analytical angular integration with the
leading-order corrections. However, we choose not to show the lengthy
expressions in this work.\footnote{In the provided subroutine for the
  2D integrals, an approximate treatment of the form factor has been
  added. However, since we only consider the leading-order term, the
  resulting opacities may not be reliable for $E_\nu \gtrsim$ 100 MeV.}
Instead, we treat the form factor effects exactly in our 4D integral
approach.

\subsection{inverse neutron decay}  

At the conditions relevant to SN matter, neutron is more massive and has a
higher mean field potential than proton. Therefore, the only allowed
decay/inverse decay is $n \rightleftarrows \bar{\nu}_e + p +
e^-$. Despite that, we still choose to consider a general process as,
$\nu_1 + N_2 + l_3 \to N_4$, where same notations are taken as in
Sec. \ref{sec:v_ab}. Note that the role of inverse neutron decay on
supernova neutrinos has recently been explored \cite{Fischer:2020kdt}.

The matrix element for the inverse decay, $\nu_1 + N_2 + l_3 \to N_4$, can be simply obtained from that for the capture, $\nu_1 + N_2 \to l_3 + N_4$, with $\nu_1, l_3$ and $N_{2,4}$ the same particle species as in the inverse decay,  
 \begin{equation}
|\mathcal{M}|^2_\text{inverse}(p_1,p_2,p_3,p_4) = |\mathcal{M}|^2_\text{capture}(p_1,p_2,-p_3,p_4).
\end{equation} 

Similarly to captures, we can perform both the 4D and the 2D
integrations. For the 4D integrals, we take the same integration
variables as for the capture, i.e.,
$E_2, |\bm{q}|, \cos\theta_{q2}, \phi_{q2}$, and the opacity due to
inverse decay is given by
\begin{equation}
\begin{aligned}
\chi(E_1) =& 2\int d\phi_{q2} \int\cos\theta_{q2} \int d|{\bm q}| \int dE_2 \langle |{\cal M}|^2 \rangle  \\
& \times f_2f_3(1-f_4) \frac{|{\bm q}| |{\bm p_2}|}{16 (2\pi)^4 E_1^2 E_4^*},  
\end{aligned}
\end{equation}
with the bounds for inverse decay determined in Sec.~\ref{sec:bound_decay}.  

Integrating out all the angles, the opacity can be expressed as a 2D integral:
\begin{equation}
\begin{aligned}
 \label{eq:imfp-decay}              
 \chi(E_1) =& \frac{G^2}{4\pi^3}\frac{1}{E_1^2}\int\limits_{E_{3-}}^{E_{3+}}dE_3 \int\limits_{E_{2-}}^{E_{2+}}dE_2 f_2f_3(1-f_4) \\
 & \times \Big[\widetilde{\mathcal{A}}\widetilde I_\mathcal{A}+\widetilde{\mathcal{B}}\widetilde I_\mathcal{B}+\dots + \widetilde{\mathcal{E}}\widetilde I_\mathcal{E} + \widetilde{\mathcal{H}}\widetilde I_\mathcal{H} \\
 & +(\widetilde{\mathcal{F}} \widetilde I_\mathcal{F}+\widetilde{\mathcal{F}}^{PP} \widetilde I_\mathcal{F}^{PP}) + (\widetilde{\mathcal{J}} \widetilde I_\mathcal{J} + \widetilde{\mathcal{J}}^{AP} \widetilde I_\mathcal{J}^{AP})  \\
 & +(\widetilde{\mathcal{K}} \widetilde I_\mathcal{K}+\widetilde{\mathcal{K}}^{AP} \widetilde I_\mathcal{K}^{AP} + \widetilde{\mathcal{K}}^{PP} \widetilde I_\mathcal{K}^{PP}) \\
 & +(\widetilde{\mathcal{L}} \widetilde I_\mathcal{L}+\widetilde{\mathcal{L}}^{AP} \widetilde I_\mathcal{L}^{AP} + \widetilde{\mathcal{L}}^{PP} \widetilde I_\mathcal{L}^{PP}) \Big],     
\end{aligned}
\end{equation}
where $\widetilde{\mathcal{X}}$, $\widetilde {\mathcal{X}}^{AP,PP}$, $\widetilde I_\mathcal{X}$, and $\widetilde I_\mathcal{X}^{AP,PP}$ can be obtained from the expressions for captures [see Eq.~(\ref{eq:imfp-I})] with the replacement  
\begin{equation}
\begin{aligned}
& \{ \widetilde{\mathcal{X}}, \widetilde{\mathcal{X}}^{AP,PP} \} = \{ \mathcal{X}, \mathcal{X}^{AP,PP} \}\Big|_{E_3 \to -E_3},  \\ 
& \{ \widetilde I_\mathcal{X}, \widetilde I_\mathcal{X}^{AP,PP} \} = -\{ I_\mathcal{X}, I_\mathcal{X}^{AP,PP} \}\Big|_{E_3 \to -E_3}.
\end{aligned}
\end{equation}
The threshold of $E_1$ and the integration bounds, $E_{2,\pm}$ and $E_{3,\pm}$, can be derived following the discussions in Sec.~\ref{sec:bound_decay}. 

\section{Results and discussion}
\label{sec:results} 

Based on the formulae presented above, we can study the neutrino
opacities from different processes. To proceed, we firstly show the
consistency between the opacities of semileptonic processes from our
2D and 4D integrals, and then discuss the effects of weak magnetism,
pseudoscalar coupling term as well the $q^2$-dependent form
factors. We also compare the different leptonic and semileptonic processes at
conditions relevant to CCSN matter.

For our studies, we take the profiles from a 2D simulation
\cite{Bollig:2017,garching} for a nonrotating 20~M$_\odot$ progenitor star
\citep{Woosley.Heger:2007} based on the Lattimer-Swesty EOS (LS200)
\citep{Lattimer.Swesty:1991}, where the relevant muonic reactions have
been implemented.\footnote{Note that the CC semileptonic processes of
  $\nu_\mu$ in \cite{Bollig:2017} were only approximately treated, not
  fully consistent with the formalisms presented in our work.} The
chemical potentials of all particles and the interaction potential of
nucleons required to obtain neutrino opacities are obtained
consistently with the same EOS.

To better demonstrate the role of weak muonic processes on
muonization, we take two conditions at $r\simeq13.6$ km (condition A)
and at 20 km (condition B) at 0.4 s after core bounce, with
$\rho \simeq 10^{14}$~g~cm$^{-3}$ and $T\simeq 38.3$~MeV, and
$\rho\simeq 1.3\times 10^{13}$~g~cm$^{-3}$ and $T\simeq 15.2$~MeV,
respectively. The chemical potentials of all relevant particles as
well as the mean field interaction potentials of nucleons derived
based on LS200 are listed in Table~\ref{tab:cond}. The corresponding
lepton fractions are $Y_e=0.13$ and $Y_\mu=0.04$ for condition A, and
$Y_e=0.11$ and $Y_\mu=0.002$ for condition B.

\begin{table}[htbp] 
  \centering 
  \caption{Two conditions considered in our study. The mean field
    potentials of nucleons are derived within the Lattimer-Swesty EoS,
    and nucleons take bare masses. All quantities are in units of
    MeV. \label{tab:cond}}
\begin{ruledtabular}
\renewcommand{\arraystretch}{1.5}
\begin{tabular}{cccccccccc}    
  & T & $\mu_n$ & $\mu_p$ & $U_n$ & $U_p$ & $\mu_e$ & $\mu_\mu$ & $\mu_{\nu_e}$ & $\mu_{\nu_\mu}$ \\    
\hline
A &  38.3 & 886.0 & 800.7 & $-24.9$ & $-42.9$ & 83.3 & 64.1 & $-2.1$ & $-20.0$ \\   
B &  15.2 & 912.7 & 875.4 & $-6.1$ & $-8.9$ & 44.4 & 37.3 & 7.1 & 1.3 \\
\end{tabular}
\end{ruledtabular}
\end{table} 

\subsection{CC semileptonic processes}

The semileptonic processes with $\mu^-$ production are only relevant 
when temperature is high enough or density is high enough so that the chemical potential difference
of nucleons, $\mu_n-\mu_p$, or the interaction potential difference,
$\Delta U_{np} = U_n-U_p$, is comparable to the muon rest mass.

\subsubsection{consistency between the 2D/4D integrals}

\begin{figure} 
\centering  
\includegraphics[width=0.51\textwidth]{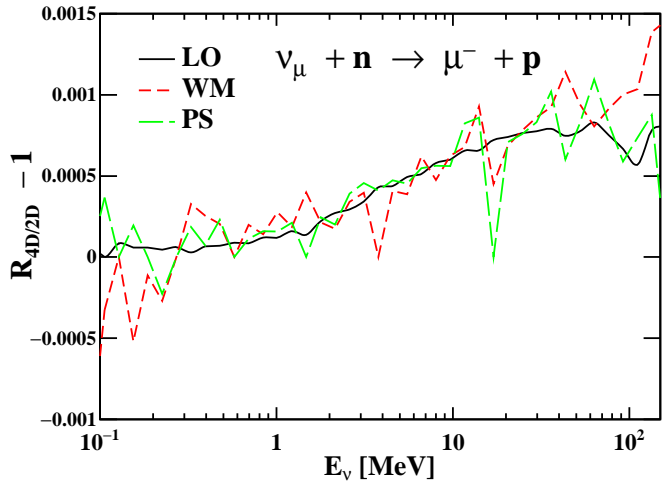}  
\caption{$R_{\rm 4D/2D}-1$ as functions of neutrino energy, $E_\nu$, with $R_{\rm 4D/2D}$ the ratio of opacity from the 4D integral to that from the 2D integral for $\nu_\mu+n\to\mu^-+p$ at condition A. Contributions from leading-order terms (LO), weak magnetism (WM), and pseudoscalar terms (PS) are considered separately. Nucleon form factors have not been considered.}                
\label{fig:R4D2D}                  
\end{figure}    

\begin{figure*}[htbp] 
\centering  
\includegraphics[width=0.49\textwidth]{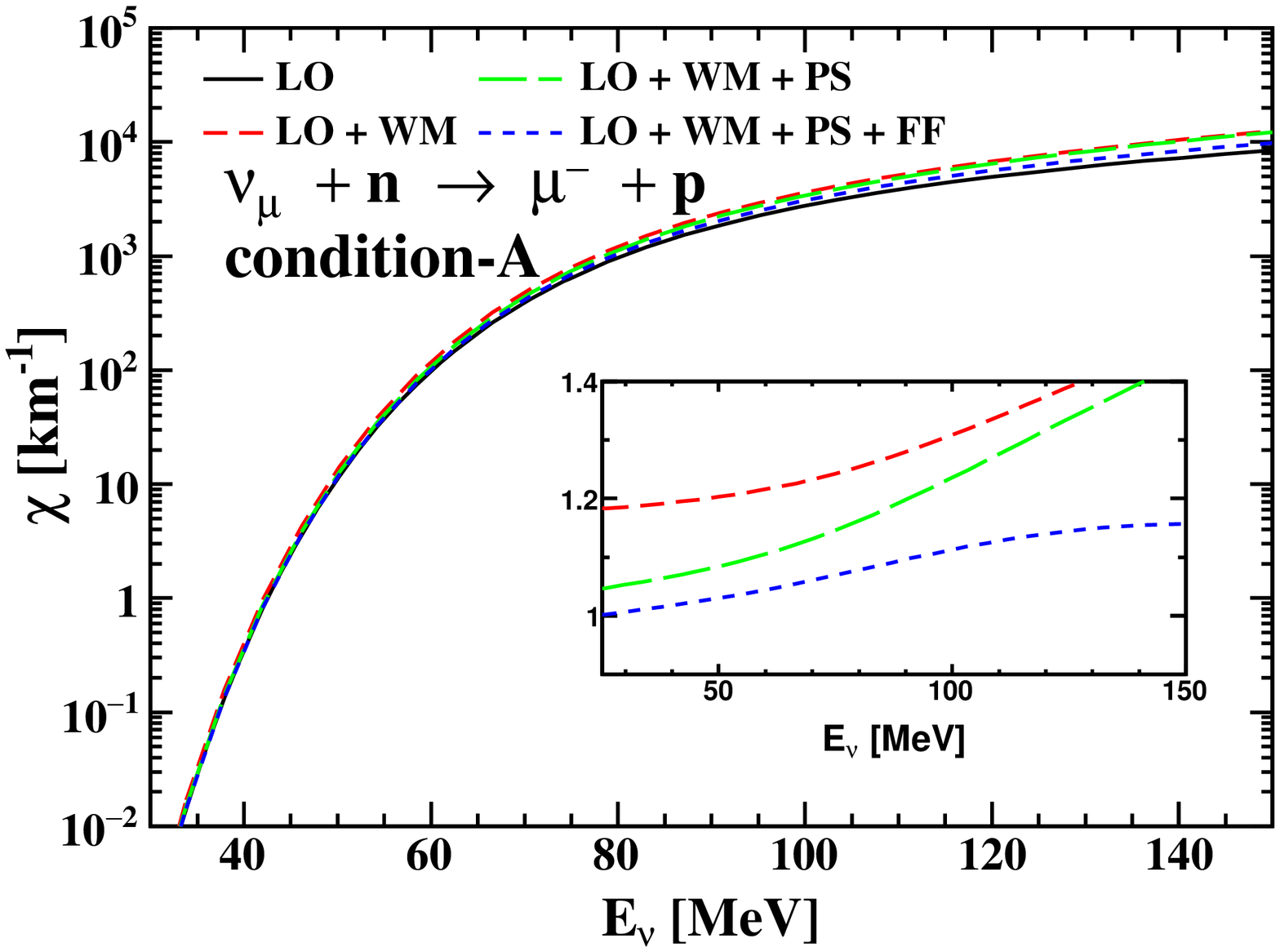}%
\includegraphics[width=0.49\textwidth]{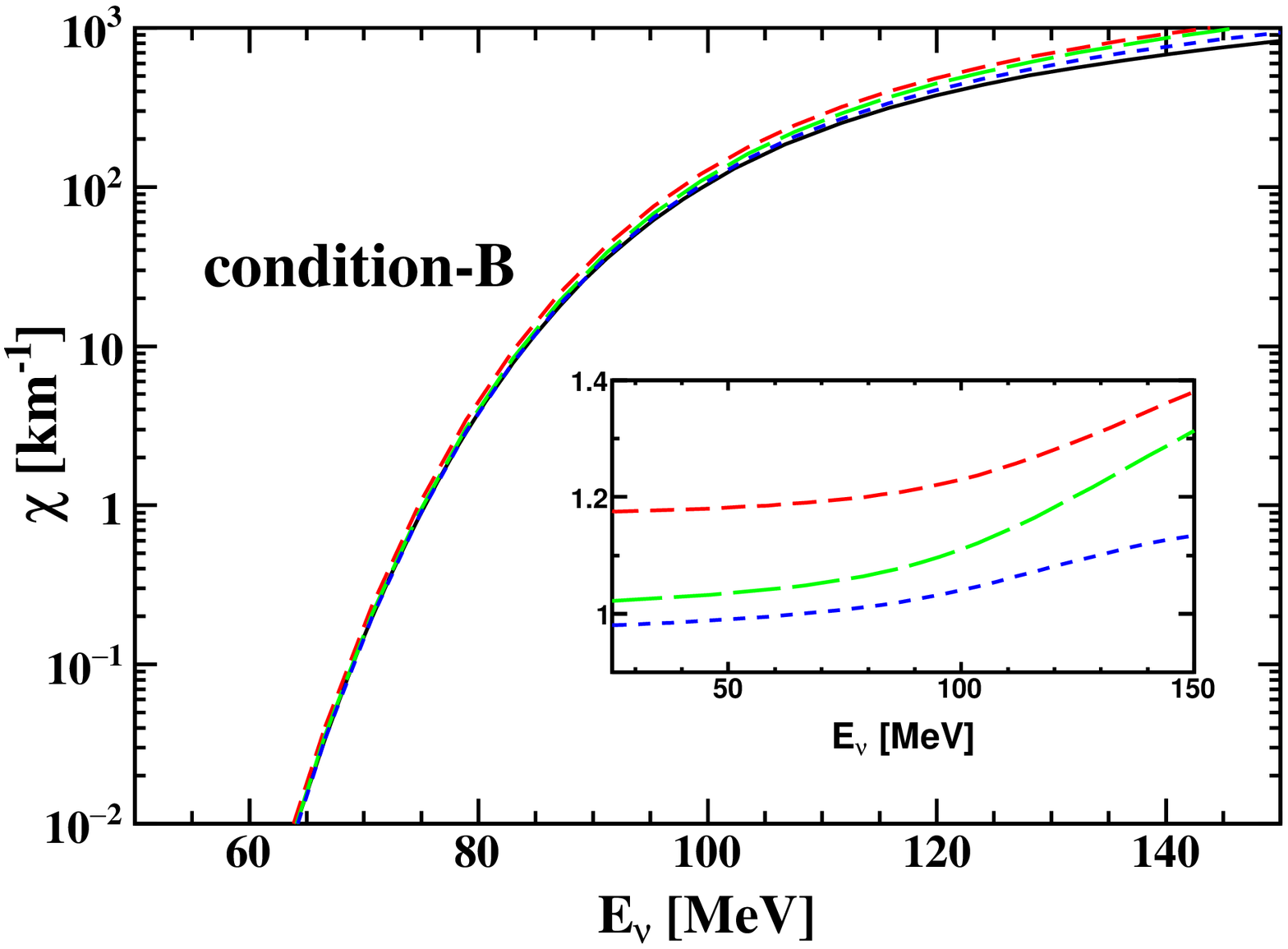}          
\caption{Neutrino opacities for $\nu_\mu+n\to \mu^-+p$ and the effects of weak magnetism (WM), pseudoscalar term (PS), and form factors (FF) at conditions A and B. The relative ratios of the opacities to that considering only the leading-order terms (LO, the black lines), i.e., terms proportional to $g_V^2, g_A^2$ and $g_Vg_A$, are shown in the inset as functions of $E_\nu$. Note that full kinematics including nucleon recoils are always considered.}                     
\label{fig:wm_pseu_form}                   
\end{figure*} 

Fig. \ref{fig:R4D2D} demonstrates the consistencies between our 4D and
2D integrals for $\nu_\mu+n\to\mu^-+p$ at condition A, where
contributions from $g_A^2, g_V^2$, and $g_Ag_V$ terms (leading-order
term, black line), from $f^2_2, g_Vf_2$, and $g_Af_2$ terms (weak
magnetism, red line), and from $g_P^2$ and $g_Ag_P$ terms
(pseudoscalar term, green line) are considered separately. The nucleon
form factors have not been considered.  As shown in the figure, the
two different methods lead to consistent opacities within $\sim$
0.1\%. The fluctuations are arisen from the Monte Carlo sampling in
the 4D integrals. Although not shown here, we have checked such
consistency also applies to the general conditions encountered in SN
simulations. Typically, it is more efficient to use the 2D integrals
than the 4D integrals. However, the 4D integral can provide a more
straightforward calculation of the opacities without any tedious
analytical derivations, and a robust check of the 2D results. Besides,
the form factor effect can be included exactly in our 4D integrals,
while the 2D integrals only treat it approximately. Without otherwise
stated, we use results from the 4D integrals for the discussions
below.

\subsubsection{effects of weak magnetism, pseudoscalar, and form factors}

\begin{figure*}[htbp] 
\centering  
\includegraphics[width=0.49\textwidth]{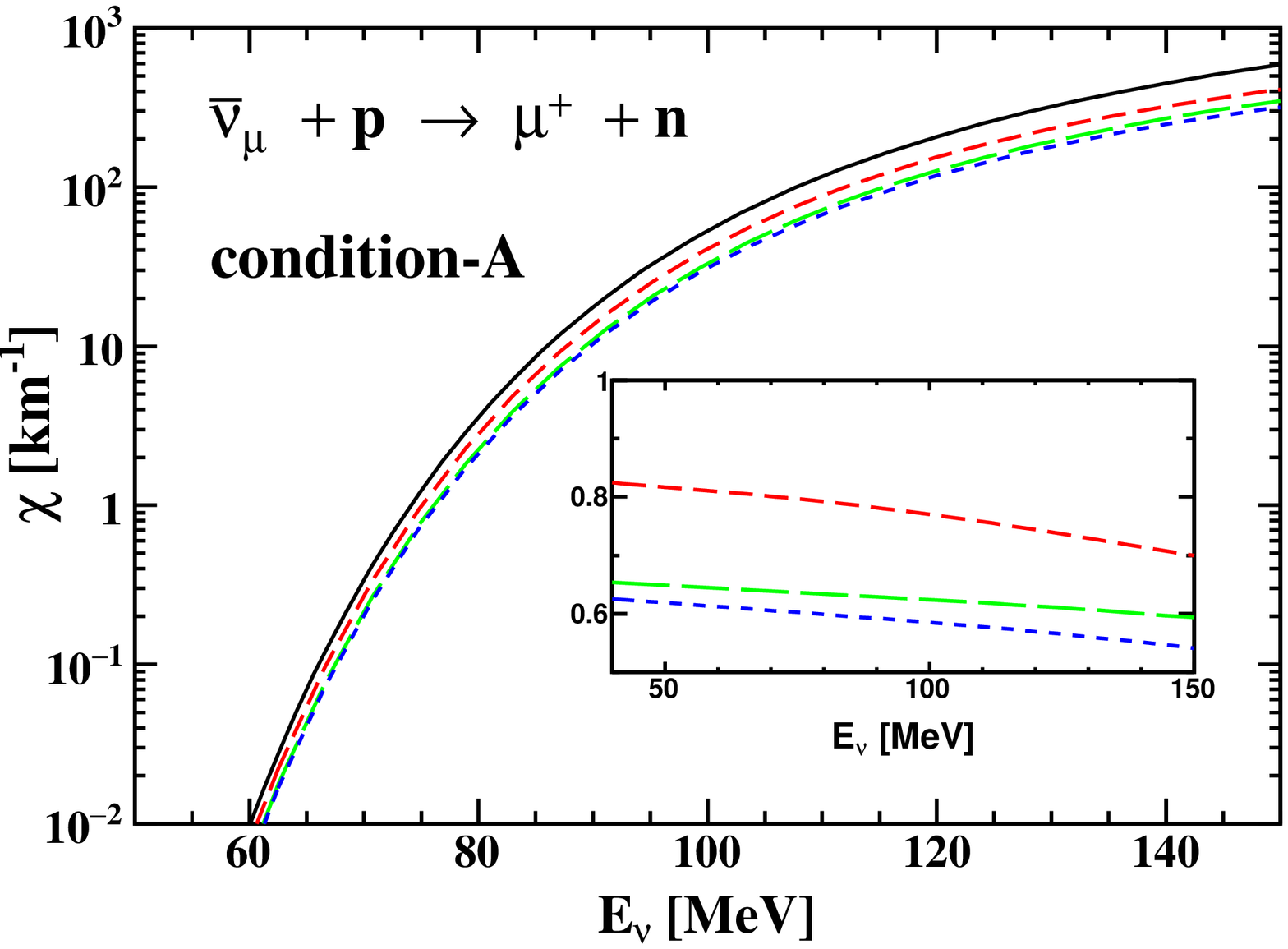}%
\includegraphics[width=0.49\textwidth]{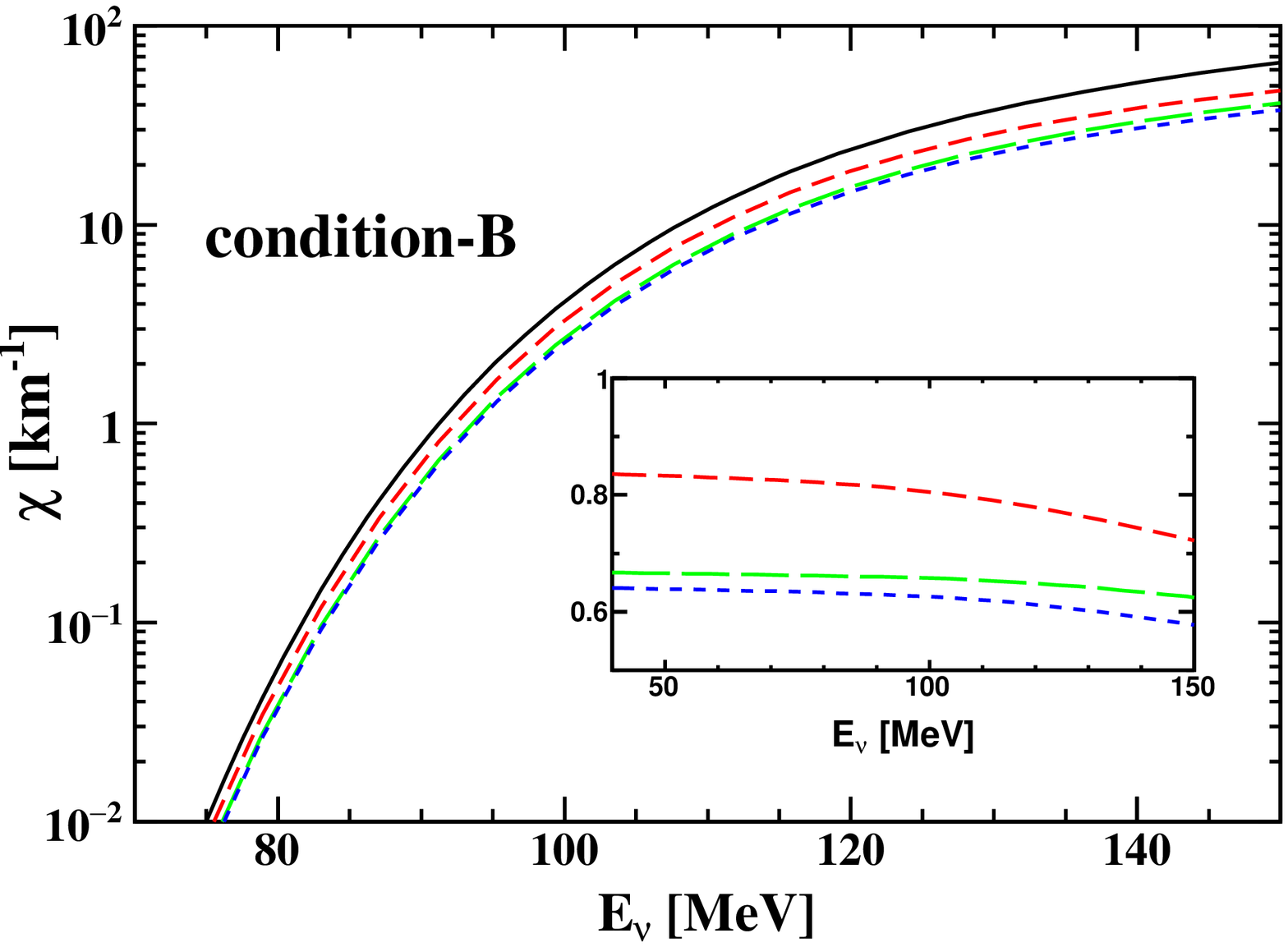}          
\caption{Same as Fig.~\ref{fig:wm_pseu_form} but for $\bar\nu_\mu+p\to \mu^++n$. 
}                    
\label{fig:anti_wm_pseu_form}                   
\end{figure*} 

Fig. \ref{fig:wm_pseu_form} compares the effects of weak magnetism (WM),
pseudoscalar term (PS), and form factors (FF) on the neutrino opacities for
$\nu_\mu+n\to \mu^-+p$ at conditions A and B. As shown in the figure,
weak magnetism enhances the absorption rates of $\nu_\mu$ by $\sim$
20\% at $E_\nu=20$ MeV and by 40\%--50\% at $E_\nu=150$ MeV, which is
consistent qualitatively with the studies in
\cite{Horowitz:2002}. However, the correction factor due to weak
magnetism introduced in \cite{Horowitz:2002} did not consider the
finite charged lepton mass and the mass/potential shifts of nucleons,
and in general can not reproduce well our weak magnetism. One can in
principle extend the analytical studies of \cite{Horowitz:2002}, which
assumes an initial neutron at rest and ignores the final state
blocking, to incorporate the mean field effects as well as a finite
muon mass. However, those approximations are not justified at
the high density and temperature conditions considered in this work.

Including the pseudoscalar term suppresses the opacities for neutrino
absorption on neutron since the interference term
$\langle|{\cal M}|^2\rangle_{AP}$ is negative, which dominates over
the positive term $\langle|{\cal M}|^2\rangle_{PP}$. The pseudoscalar
term has a large impact at small $E_\nu$, which then decreases with
increasing $E_\nu$. Such behaviour can be understood as
follows. Neglecting the mean field effects on nucleon masses and
$\Delta U_{np}^2$ terms, the matrix element
$\langle|{\cal M}|^2\rangle_{AP}$ shown in Eq.~(\ref{eq:MsqAP}) is
proportional to $[-(p_1 \cdot p_3)+2E_\nu\Delta U_{np}]m_\mu^2$ with
$(p_1 \cdot p_3) \sim m_\mu^2$. As $E_\nu$ increases, the second term
becomes more relevant and cancels the first term. We also note that
the pseudoscalar term has negligible effects on $\nu_e$ reactions as
it scales with $m_l^2$. Differently from weak magnetism, the
pseudoscalar term also suppresses the opacity for antineutrino
absorption.

The reduction of opacities due to the $q^{2}$-dependent form factors
is easy to understand since $q^{2}$ is always negative [see
Eq.~(\ref{eq:form-factor})]. As $|q^{2}|$ increases with $E_\nu$, the
effects will become more significant. For $\nu_\mu$ absorption, the
enhancement due to weak magnetism, of (20--30)\% at $E_\nu=100$ MeV,
is largely cancelled by the effects of pseudoscalar term and form
factor. The net effect, compared to the leading-order results, is to
enhance the opacities by only about 10\% and 4\% at $E_\nu=100$ MeV
for conditions A and B, respectively. We have also checked that such
cancellation is quite robust for conditions relevant to supernova
matter. For the regions of interest, all effects combined enhance the
opacities of $\nu_\mu$ of energies around 100 MeV by $\lesssim$ 10\%
compared to the leading-order results.

As shown in Fig.~\ref{fig:anti_wm_pseu_form}, the opacities of
$\bar\nu_\mu$ from $\bar\nu_\mu+p\to \mu^+ + n$ are suppressed by
$\sim$ 40\% for all the relevant $E_\nu$ at both conditions, as all
effects contribute negatively. Since protons are in lower energy
states compared to neutrons, the rates of $\bar\nu_\mu$ absorption on
protons are about 30 times slower than those of $\nu_\mu$.

\begin{figure}[htbp] 
\centering  
\includegraphics[width=0.49\textwidth]{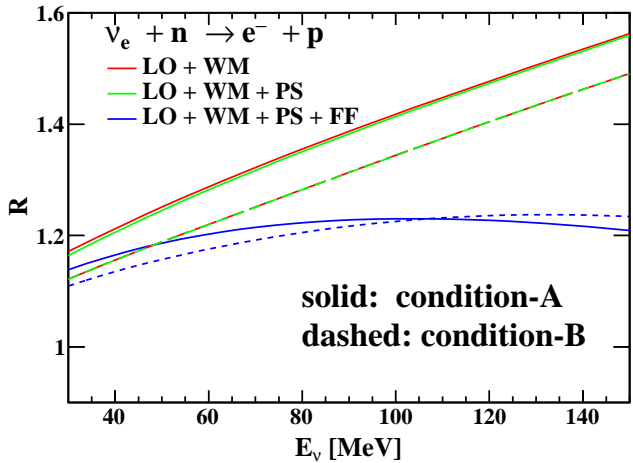}  
\caption{Ratios of $\nu_e$ opacities for $\nu_e+n\to e^-+p$ including weak magnetism, pseudoscalar term, and form factor effects to the leading-order results.}                 
\label{fig:nu_e_CC}                  
\end{figure}

The transport of $\nu_e$ is typically dominated by the CC absorption
on neutrons in hot PNS. Therefore, an accurate description of this
process is important. We show in Fig.~\ref{fig:nu_e_CC} the effects of
different treatments of the hadronic current for $\nu_e$. As mentioned
above, the pseudoscalar term has negligible effects. However, the
effect of form factors is non-negligible at high $E_\nu$. It can
suppress the $\nu_e$ absorption rates by (10--20)\% for
$E_\nu \gtrsim 100$ MeV. The net effect incorporating all corrections
is to enhance the leading-order results by $\sim 20$\% at conditions A
and B.

\subsection{muonization from leptonic and semileptonic processes}

\begin{figure*}[htbp] 
  \centering
  \includegraphics[width=0.49\textwidth]{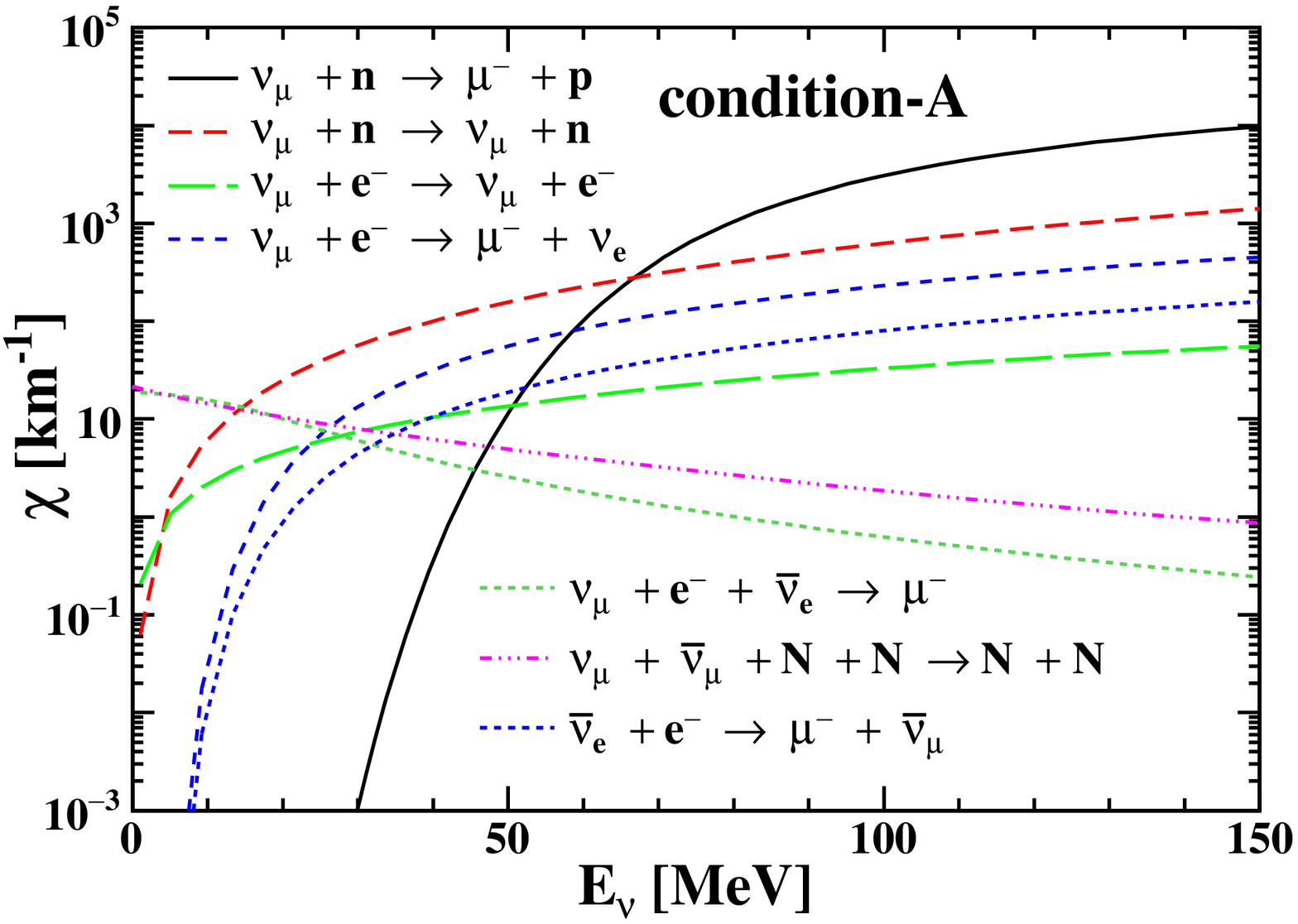}%
  \includegraphics[width=0.49\textwidth]{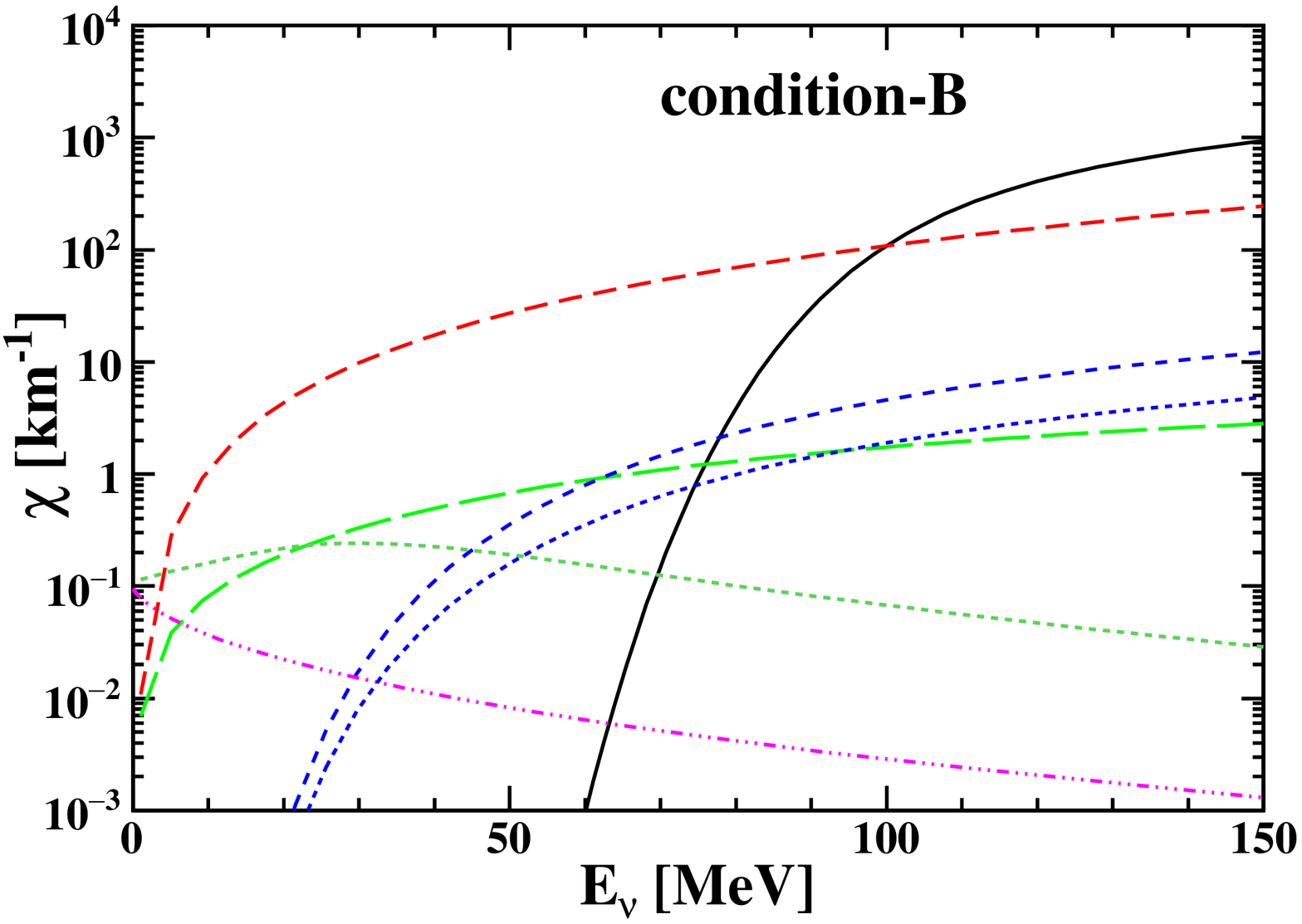}
\caption{Comparison of $\nu_\mu$ opacities from different weak muonic processes. The left and right plots are for conditions A and B, respectively. We also include $\bar\nu_e+e^- \to \mu^- + \bar\nu_\mu$ to show its relevance in producing $\mu^-$.}                      
\label{fig:numu_opa}                    
\end{figure*}

Fig. \ref{fig:numu_opa} compares the opacities of different weak muonic processes at conditions A and B.
There are mainly four weak processes contributing to $\mu^-$ production:
$\nu_\mu+n\to \mu^-+p$, $\nu_\mu+e^-\to \mu^-+\nu_e$, $\bar\nu_e+e^-\to\bar\nu_\mu+\mu^-$, and $\nu_\mu+e^-+\bar\nu_e \to \mu^-$.
The semileptonic processes always dominate at high $E_\nu$ due to larger matrix elements. At low $E_\nu$, purely leptonic reactions between $\nu_\mu$ and electrons become more important for $\mu^-$ production due to a larger physically allowed phase space. Especially, inverse muon decay (blue short-dashed lines) allows neutrinos with energies below 10--20 MeV to contribute.
With a Fermi-Dirac distribution, the averaged energies of neutrinos are about $3.15 \times T$.
For condition A with higher temperature and density ($\langle E_\nu
\rangle \sim 100$ MeV), the semileptonic processes are the main
production channel for $\mu^\pm$, while at outer region with lower
temperature (condition B with $\langle E_\nu \rangle \sim 50$ MeV),
leptonic processes are more relevant. Compared to $\mu^-$, the rates
of $\mu^+$ production via semileptonic or leptonic processes are
slower by a factor of $\sim$ 30, leading to a gradual buildup of
$\mu^-$ excess over $\mu^+$. Note that although the electromagnetic
processes like $e^-+e^+ \to \mu^-+\mu^+$ and $\gamma+\gamma \to
\mu^-+\mu^+$ are much faster than the weak processes, they do not
contribute directly to muonization since $\mu^\pm$ are always produced
in pairs. Considering a high electron chemical potential (with
$\mu_e\approx 83$ MeV and $\approx 44$ MeV for conditions A and B,
respectively), $\mu^-$ decay rate is lower than that of $\mu^+$ due to
final state $e^-$ blocking,
which could also play a role in muonization.

The electromagnetic processes should act fast enough to keep muons in
local thermal equilibrium. Due to muon number conservation,
muonization is further determined by $\nu_\mu$ and $\bar\nu_\mu$
transport, and specifically, their difference. As shown in
Fig.~\ref{fig:numu_opa}, neutrino opacities are dominated by
semileptonic processes with nucleons. For NC neutrino-nucleon
scattering (red dashed lines), we take the elastic approximation as in
Ref.~\cite{Bruenn:1985} and only consider contributions from
scattering with neutrons. Inclusion of weak magnetism leads to a
higher $\nu_\mu$ opacity from NC scattering with nucleons compared to
that of $\bar\nu_\mu$ \cite{Horowitz:2002}, and thus affects
muonization \cite{Keil.Raffelt.Janka:2003}.  The CC semileptonic
reaction rates of $\nu_\mu$ are even higher than those of the NC ones
at high $E_\nu$, making $\nu_\mu$ more easily trapped. In addition to
these semileptonic processes, inverse muon decay can be important
opacity sources for low energy $\nu_\mu$. We also include neutrino
pair annihilation on two nucleons (also called inverse bremsstrahlung)
based on the T-matrix formalism from Ref. \cite{Guo:2019cvs}, and find
that it can contribute with a comparable amount as inverse muon decay
at low $E_\nu$. Scattering or absorption of $\nu_\mu$ on electrons is
relevant at intermediate $E_\nu$, and can play a role in determining
the neutrino spectra via efficient energy exchange. Scattering with
$e^+$ and $\mu^-$ also contribute to $\nu_\mu$ opacities and can be
included similarly.  Besides, $\nu_\mu \bar\nu_\mu$ pair annihilation
to $e^\pm$ or other neutrino pairs and $\nu_\mu$-$\nu_\alpha$
scattering are subdominant for most of the relevant conditions and are
not shown.\footnote{Note that the inverse process, i.e, $e^\pm$
  annihilation to neutrino pair, is important for the thermal
  production of neutrinos, especially for the heavy flavours.}
 It should be mentioned that at high temperature and close to saturation density $\nu_\mu + \pi^- \to \mu^-$ is kinematically allowed due a strong attractive potential for pions, and could dominate the opacity for $E_\nu \lesssim 20$ MeV \cite{Fore:2019wib}.          

\subsection{inverse neutron decay and inverse muon decay as opacity source for $\bar\nu_e$}    

\begin{figure*}[htbp] 
\centering  
\includegraphics[width=0.49\textwidth]{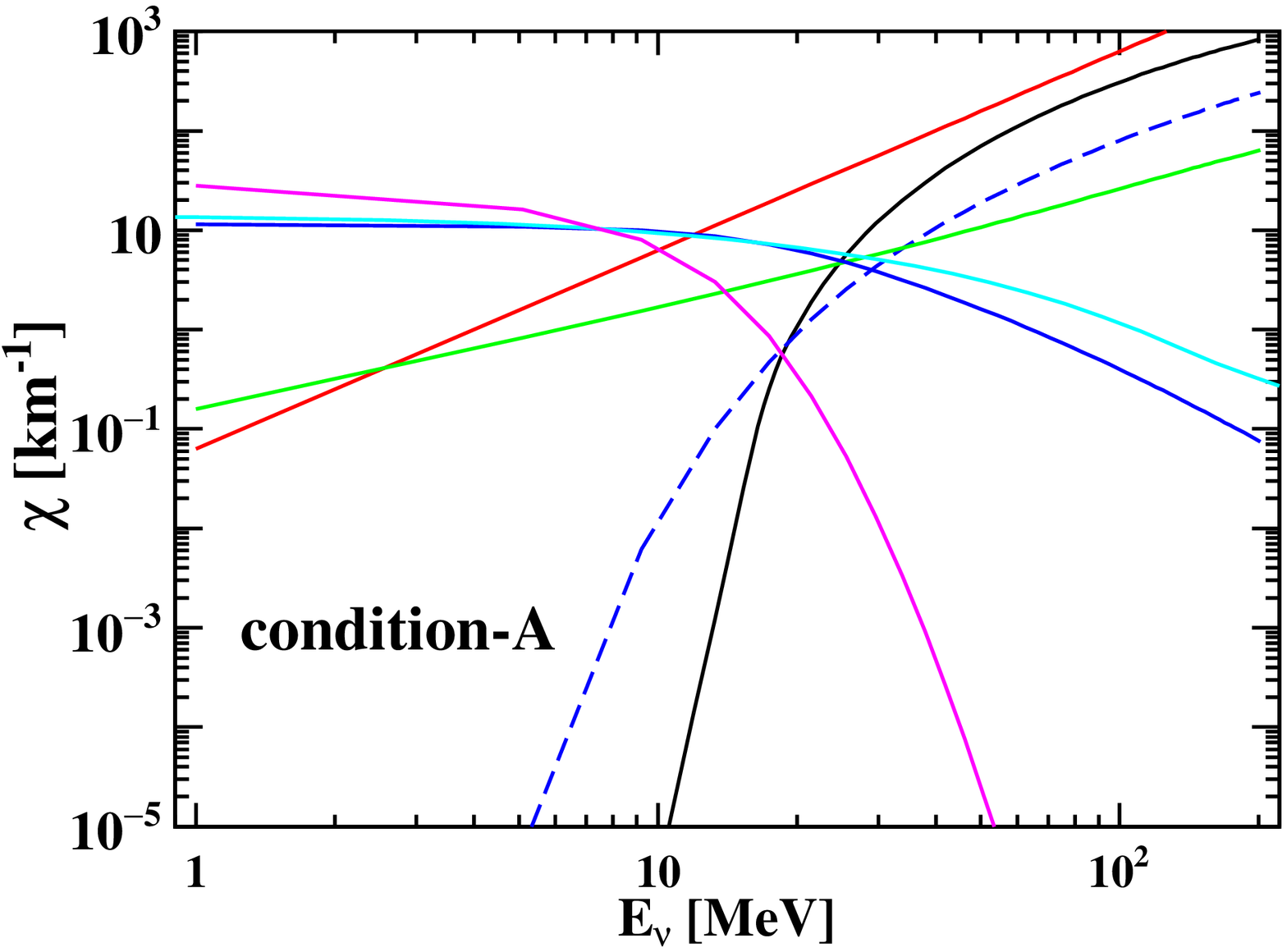}%
\includegraphics[width=0.49\textwidth]{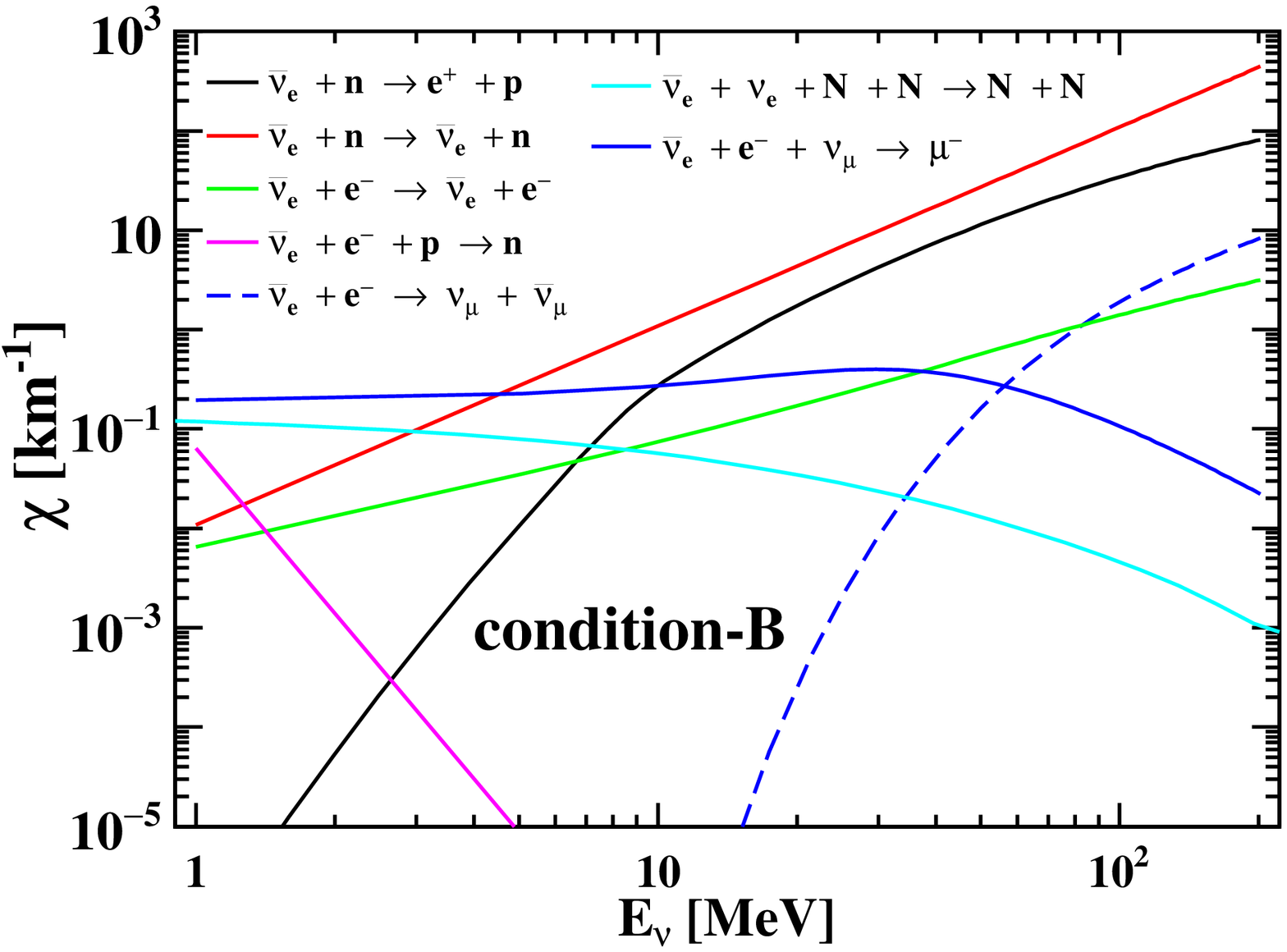}          
\caption{Comparison of $\bar\nu_e$ opacities from different weak processes. The left and right plots are for conditions A and B, respectively.}                      
\label{fig:nueb_opa}                    
\end{figure*}

As mentioned above, all the formulae presented in this work can apply
equally to $\nu_e$ and $\bar\nu_e$. The opacity of $\bar\nu_e$ is
dominated by NC scattering on neutrons followed by CC absorption on
protons. With increasing density, the energy difference between
neutrons and protons increases and hence the absorption rate at low $E_{\bar{\nu}_e}$ is strongly suppressed.
On the other hand,
inverse processes including inverse neutron decay, inverse muon decay
as well as inverse bremsstrahlung could be important opacity
sources. In Fig.~\ref{fig:nueb_opa}, we compare opacities of
$\bar\nu_e$ from all relevant processes. At condition A, all the
inverse processes dominate the $\bar\nu_e$ opacity at low energies
below $\sim 10$ MeV. At condition B, the relevance of these inverse
processes become less significant as the phase space shrinks with
temperature. The phase space for inverse neutron decay rate is further
suppressed by a smaller $\Delta U_{np}$. The role of inverse neutron
decay on supernova neutrinos has recently been studied
\cite{Fischer:2020kdt}. Note that all the
corrections in the hadronic
current, including weak magnetism, pseudoscalar term, and form factor effect, can be neglected for inverse neutron decay as they are only
important at high $E_\nu$.

\section{subroutines for neutrino absorption on nucleons} 

We provide subroutines for calculating CC neutrino-nucleon
opacities for CCSN simulations at \url{https://git.gsi.de/nucastro_archive/supernovamuonicrates}. Specifically, two
different schemes discussed in this work, one invoking a 2D integral
and the other using a 4D Monte Carlo integration, have been applied to
calculate the opacities. For both subroutines, interaction terms
including weak magnetism, pseudoscalar term as well as form factor
effects can be included or turned off. As mentioned, the 2D subroutine
only treats the leading-order terms of the form factor effects, and
may give wrong opacities for $E_\nu$ above $\sim$100 MeV, while the 4D
subroutine provides an exact description. In principle, both
subroutines can be called in-flight for neutrino transport in CCSN
simulations. The 2D subroutine accounting for weak magnetism using
32-grid-point Gauss quadratures for each dimension has been shown to
be efficient and stable in symmetric CCSN simulations
\cite{Fischer:2020kdt}.\footnote{We have checked that the 2D
  subroutine using 32 grids for each dimension are enough to reach an
  accuracy of 5\%.} The 4D subroutine, on the other hand, could be
inefficient. Using less grid points or Monte Carlo points may provide
ways to trade-off accuracy for computational efficiency for both
subroutines. Another possibility is to first tabulate neutrino opacities consistent with the nuclear EoS.

\section{Summary}    
\label{sec:summary}           

In this work we derive analytical formulae for CC reactions involving $\nu_\mu$/$\bar\nu_\mu$ or $\mu^\pm$ with full relativistic kinematics in the mean field approximation. For the semileptonic reactions, we present an accurate treatment of the hadronic weak current including weak magnetism, pseudoscalar term, and form factor effects. Although weak magnetism increases $\nu_\mu$ reaction rate, the pseudoscalar term and the nucleon form factors tend to reduce it. For antineutrinos, on the other hand, all these corrections suppress the opacities. The formulae presented in this work can also be applied to $\nu_e$ ($\bar\nu_e$) reactions. Although the pseudoscalar term has negligible effect, the nucleon form factors can reduce the $\nu_e$ rate and cancel the enhancement effect due to weak magnetism as for $\nu_\mu$. 

The muonic semileptonic process dominates the $\nu_\mu$ opacity at $E_\nu \gtrsim 60$--100 MeV and plays essential roles in $\mu^-$ production and muonization shortly after core bounce. Purely leptonic processes involving $\nu_\mu$ or $\mu^-$ have also been studied and compared to the semileptonic ones. Scattering and absorption on electrons are major inelastic channels for low and intermediate energy $\nu_\mu$. At densities higher than $\sim 10^{13}~\mathrm{g~cm^{-3}}$ as explored in this work, inverse muon decay becomes the dominant inelastic reaction at $E_\nu \lesssim 20$ MeV. As density further increases, inverse muon decay as well as inverse bremsstrahlung can contribute predominantly to $\nu_\mu$ opacities of energy below $\sim 10$ MeV.         

The opacity for $\bar\nu_e$ absorption on proton is suppressed due to a relatively low proton density or a larger energy difference between neutrons and protons at high densities. Similarly to $\nu_\mu$, inverse muon decay and inverse bremsstrahlung are also shown to be important reactions for low energy $\bar\nu_e$ at densities higher than $\sim 10^{13}~\mathrm{g~cm^{-3}}$. As the energy difference between neutrons and protons in nuclear medium increases, inverse neutron decay, with a larger phase space available, also acts as an important opacity source for low energy $\bar\nu_e$. 

To study the impact of different weak processes, especially muonic reactions, we have considered two specific conditions encountered in PNS $\sim$ 0.4 s post core bounce, with densities $\gtrsim 10^{13}~\rm{g~cm^{-3}}$. Although they may not contribute directly in the decoupling region with low densities, the early muonization and the sequent transport of neutrinos could have a non-negligible impact. In particular, considering that low energy neutrinos decouple in deeper regions, these inverse processes could be more interesting. As the PNS contracts and cools down, the effective neutrinosphere will move to high density regions, making the new reactions studied in our work more relevant.
In a companion paper of us to be submitted, the impact of these weak muonic processes on the evolution of PNS and neutrino emissions has also been quantified. We also provide subroutines for computing the CC neutrino-nucleon reaction rates, which can be applied in CCSN simulations.

\begin{acknowledgments}
  We thank Robert Bollig and Thomas Janka for providing profiles of a 2D supernova model from
Ref. \cite{Bollig:2017} for this work. GG acknowledges support from the Academia Sinica by Grant
  No.~AS-CDA-109-M11. GMP acknowledges the support of the
  Deutsche Forschungsgemeinschaft (DFG, German Research Foundation) --
  Project-ID 279384907 -- SFB 1245 ``Nuclei: From Fundamental
  Interactions to Structure and Stars''. TF acknowledges  support from the Polish National Science Center (NCN) under grant numbers 2016/23/B/ST2/00720 and 2019/33/B/ST9/03059. 
\end{acknowledgments}

\appendix

\begin{widetext}

\section{amplitudes and coefficients for leptonic reactions}
\label{sec:append_lep}

For all the leptonic processes (a)-(e) listed in Table~\ref{tab:mu_reaction}, the coefficients $\lambda_i$ in Eq.~\eqref{eq:kernel_lep} for neutrino scattering on charged leptons and $\lambda_{Ai}$ in Eq.~\eqref{eq:msq_lep_inv} for inverse muon decay are given by:
\begin{align}
& \lambda_1^{\rm (a)} = 16G_{\rm F}^2(1-4s_W^2+4s_W^4),\;\; 
\lambda_2^{\rm (a)} = 64G_{\rm F}^2 s_W^4,\;\; 
\lambda_3^{\rm (a)} = 16G_{\rm F}^2 m_e^2 (2s_W^2-4s_W^4), \\
& \lambda_1^{\rm (b)} = 64G_{\rm F}^2 s_W^4, \;\;
\lambda_2^{\rm (b)} = 16G_{\rm F}^2 (1-4s_W^2+4s_W^4), \;\; 
\lambda_3^{\rm (b)} = 16G_{\rm F}^2 m_e^2 (2s_W^2-4s_W^4), \\
& \lambda_1^{\rm (c)} = 64G_{\rm F}^2 , \;\; \lambda_2^{\rm (c)} =0,\;\;  \lambda_3^{\rm (c)} =0, \\
& \lambda_1^{\rm (d)} = 0,\;\; \lambda_2^{\rm (d)}=64G^2_{\rm F}, \;\; \lambda_3^{\rm (d)} =0,  \\ 
& \lambda_{A1}^{\rm (e)} = 64G_{\rm F}^2 ,\;\; \lambda_{A2}^{\rm (e)}=0, \;\; \lambda_{A3}^{\rm (e)} =0.
\end{align} 

The coefficients $A_i$, $B_i$, and $C_i$ in Eq.~(\ref{eq:R_i}) are universal for all scattering processes and are given by 
\begin{subequations}
\begin{equation}
\begin{aligned}
A_1 =& E_1 E_3 (1-\mu)^2 [E_1^2 + E_1E_3(3+\mu)+E_3^2], 
\end{aligned}
\end{equation} 
\begin{equation}
\begin{aligned}
B_1 =& E_1^2 E_3 (1-\mu)^2 [2E_1^2 + E_1E_3(3-\mu) -E_3^2(1+3\mu)]  
 + Q(1-\mu)[  E_1^3+E_1^2E_3(2+\mu) -E_1E_3^2(2+\mu)-E_3^3],  
\end{aligned}
\end{equation} 
\begin{equation}
\begin{aligned}
C_1 =& E_1^3 E_3 (1-\mu)^2 \Big[E_1^2 - 2E_1E_3\mu+E_3^2\Big(-\frac{1}{2}+\frac{3}{2}\mu^2\Big)\Big]  
 + QE_1(1-\mu)[ E_1^3-E_1^2E_3\mu+E_1E_3^2(-2+\mu^2)+E_3^3\mu] \\
& + Q^2 \Big[E_1^2\mu-E_1E_3\Big(\frac{3}{2}+\frac{1}{2}\mu^2\Big)+E_3^2\mu\Big] 
 +\frac{1}{2}E_1E_3(1-\mu^2)\Delta^2 m_2^2,   
\end{aligned}
\end{equation}
\begin{equation}
\begin{aligned}
\{A_2, B_2, C_2\}=& \{A_1, B_1, C_1\}\Big\vert_{E_1 \leftrightarrow -E_3},  
\end{aligned}
\end{equation} 
\begin{equation}
A_3=B_3=0,\;\; C_3=(1-\mu)\Delta^4, 
\end{equation} 
\end{subequations}
where $\Delta=\sqrt{E_1^2-2E_1E_3\mu+E_3^2}$,  $Q=(m_4^2-m_2^2)/2$, and the coefficients $A_{Ai}, B_{Ai}, C_{Ai}$ for inverse decay in Eq.~(\ref{eq:R_Ai}) are simply 
\begin{align}
\{A_{Ai}, B_{Ai}, C_{Ai}\}=& \{A_i, B_i, C_i\}\Big\vert_{E_3 \rightarrow -E_3}.    
\end{align}

\section{Matrix elemtents and coefficients for semileptonic reactions}
\label{sec:append_semi}

\subsection{matrix elements} 
Matrix elements of neutrino absorption on nucleon, $\nu_l (p_1) + N_2 (p_2) \to l^- (p_3) + N_4 (p_4)$, as shown in Eq.~\eqref{eq:matrixedecomp}, are 
\begin{subequations}    
\label{eq:semi_amp}
\begin{equation}
\langle|{\cal M}|^2\rangle_{VV} = 16G^2 G_V^2 \Big[ (p_1\cdot p_2^*)(p_3 \cdot p_4^*)+ (p_1\cdot p_4^*)(p_3 \cdot p_2^*)  \\
 - m_2^* m_4^*(p_1 \cdot p_3)\Big],   
\end{equation}
\begin{equation}
\langle|{\cal M}|^2\rangle_{VA} = 32G^2 G_V G_A \Big[ (p_1\cdot p_2^*)(p_3 \cdot p_4^*) - (p_1\cdot p_4^*)(p_3 \cdot p_2^*)\Big],   
\end{equation}
\begin{equation}
|{\cal M}|^2_{AA} = 16G^2 G_A^2 \Big[ (p_1\cdot p_2^*)(p_3 \cdot p_4^*)+ (p_1\cdot p_4^*)(p_3 \cdot p_2^*) + m_2^* m_4^*(p_1 \cdot p_3)\Big],   
\end{equation}
\begin{equation}
\begin{aligned}
\langle|{\cal M}|^2\rangle_{VF} = \frac{16G^2 G_VF_2}{2M_N}\Big\{ & \Big[(p_1\cdot p_2^*)m_4^* - (p_1\cdot p_4^*)m_2^* \Big](p_3 \cdot q^*)  
+  \Big[(p_3\cdot p_2^*)m_4^*-(p_3\cdot p_4^*)m_2^*\Big](p_1 \cdot q^*)   \\
+ & \Big[(q^* \cdot p_2^* )m_4^* - (q^* \cdot p_4^*)m_2^*\Big](p_1 \cdot p_3) \Big\},  
\end{aligned} 
\end{equation} 
\begin{equation}
\langle|{\cal M}|^2\rangle_{AF} = \frac{16G^2 G_AF_2}{M_N}\Big\{ \Big[(p_1\cdot p_2^*)m_4^* + (p_1\cdot p_4^*)m_2^* \Big](p_3 \cdot q^*)  
-  \Big[(p_3\cdot p_2^*)m_4^*+(p_3\cdot p_4^*)m_2^*\Big](p_1 \cdot q^*) \Big\},   
\end{equation}
\begin{equation}
\begin{aligned}
\langle|{\cal M}|^2\rangle_{FF} = \frac{16G^2 F_2^2}{8M_N^2}\Big\{      
& 2 \Big[(p_3 \cdot p_2^*) (p_4^* \cdot q^*) + (p_3 \cdot p_4^*) (p_2^* \cdot q^*) \Big](p_1 \cdot q^*)   
+2 \Big[(p_1 \cdot p_2^*) (p_4^* \cdot q^*) + (p_1 \cdot p_4^*) (p_2^* \cdot q^*) \Big](p_3 \cdot q^*) \\  
-& 2 (p_1 \cdot q^*) (p_3 \cdot q^*) (p_2^* \cdot p_4^*)     
+ q^{*2}  \Big[ (p_1\cdot p_3) (p_2^* \cdot p_4^*) - 2( p_1 \cdot p_2^*)(p_3 \cdot p_4^*)  
  - 2 ( p_1 \cdot p_4^*)(p_3 \cdot p_2^*)\Big]    \\ 
-& m_2^* m_4^*\Big[  (p_1 \cdot p_3) q^{*2} + 2(p_1 \cdot q^*)(p_3 \cdot q^*) \Big] \Big\},   
\end{aligned}
\end{equation}
\begin{equation}
\begin{aligned}
\langle|{\cal M}|^2\rangle_{AP} = \frac{16G^2 G_AG_P}{M_N}\Big\{ & (p_1 \cdot q^*) \Big[ (p_3 \cdot p_2^*)m_4^* - (p_3 \cdot p_4^*)m_2^*\Big] 
+  (p_3 \cdot q^*) \Big[ (p_1 \cdot p_2^*)m_4^* - (p_1 \cdot p_4^*)m_2^*\Big]  \\ 
+ & (p_1 \cdot p_3) \Big[  (q^* \cdot p_4^*)m_2^* - (q^* \cdot p_2^*)m_4^*\Big] \Big\},
\end{aligned}
\label{eq:MsqAP}
\end{equation}
\begin{equation}   
\langle|{\cal M}|^2\rangle_{PP} = \frac{16G^2 G_P^2}{2M_N^2} \Big[2( p_1 \cdot q^* )(p_3 \cdot q^*) - q^{*2}(p_1\cdot p_3)\Big]  
 \Big[(p_2^* \cdot p_4^*) - m_2^* m_4^*\Big], 
\end{equation}
\end{subequations} 
where $G=G_{\rm F} V_{ud}$ with $G_{\rm F}$ the Fermi coupling constant and
$V_{ud}$ the up-down entry of the Cabibbo-Kobayashi-Maskawa matrix.

\subsection{ $I_\mathcal{X}$ and the coefficients $\mathcal{X}$ in Eq.~(\ref{eq:imfp-I})} 

We take $I_\mathcal{A}$ for example, and demonstrate how the angular integration can be solved. From the definitions in Eqs.~(\ref{eq:semi_amp_decomp}) and (\ref{eq:I_x}), $I_\mathcal{A}$ is given by
\begin{align}
 I_\mathcal{A} = & \frac{\bar p_1\bar p_2\bar p_3\bar p_4}{4\pi^2}\int d\Omega_2 d\Omega_3 d\Omega_4 \left(p_1\cdot p_2^*\right) \left(p_3\cdot p_4^*\right) \nonumber\\
 & \times \delta^{(3)}\!\left(\bm{p}_1\!+\!\bm{p}_2\!-\!\bm{p}_3\!-\!\bm{p}_4\right).
\end{align}
Define $\bm{p}_a=\bm{p}_1+\bm{p}_2$ and $\cos{\theta_2}=x_2$ with $\theta_2$ the angle between $\bm{p}_1$ and $\bm{p}_2$, we have 
\begin{align}
 \bar p_a=\sqrt{\bar p_1^2+\bar p_2^2+2\bar p_1\bar p_2x_2}, \label{eq:pa_x2}
\end{align}
with the notation $\bar p_i= |\bm{p}_i|$. The momentum delta function can be expressed as
\begin{align}
 &\delta^{(3)}\!\left(\bm{p}_1\!+\!\bm{p}_2\!-\!\bm{p}_3\!-\!\bm{p}_4\right) \nonumber\\
 & = \frac{1}{\bar p_4^2} \delta{\left(\bar p_4-\left|\bm{p}_a-\bm{p}_3\right|\right)} \delta^{(2)}{\left(\Omega_4-\Omega_{\left|\bm{p}_a-\bm{p}_3\right|}\right))}.
\end{align}
Integrating over $d\Omega_4$, we obtain  
\begin{align}
 I_\mathcal{A} = & \frac{\bar p_1\bar p_2\bar p_3}{4\pi^2\bar p_4}\int d\Omega_2 d\Omega_3 \left(E_1E_2^*-\bm{p}_1\cdot\bm{p}_2\right) \left(E_3E_4^*-\bm{p}_3\cdot\bm{p}_4\right) \nonumber\\
 & \times \delta{\left(\bar p_4-\left|\bm{p}_a-\bm{p}_3\right|\right)}.
\end{align}
The angular integral for particle `3' is given by $d\Omega_3=d\phi_3 dx_3$. One has the freedom to define $x_3 = \cos\theta_{a3}$, with $\theta_{a3}$ the angle between $\bm{p}_a$ and $\bm{p}_3$.
This allows for a variable substitution in the $\delta$-distribution
\begin{align}
\label{eq:Fix-x3}
 \delta{\left(\bar p_4-\left|\bm{p}_a-\bm{p}_3\right|\right)}=\frac{\bar p_4}{\bar p_a\bar p_3}\delta{\left(x_3-\frac{\bar p_a^2+\bar p_3^2-\bar p_4^2}{2\bar p_a\bar p_3}\right)}.
\end{align}
Now one can rewrite the momentum product $\bm{p}_3\cdot\bm{p}_4$ as
\begin{align}
 \bm{p}_3\cdot\bm{p}_4=\bm{p}_3\cdot(\bm{p}_a-\bm{p}_3)=\bar p_3\bar p_ax_3-\bar p_3^2=\frac{\bar p_a^2-\bar p_3^2-\bar p_4^2}{2}.
\end{align}
Similarly it is trivial to show that 
\begin{align}
 \bm{p}_1\cdot\bm{p}_2=\frac{\bar p_a^2-\bar p_1^2-\bar p_2^2}{2}.
\end{align}
Performing integration over $\phi_3$ and $x_3$, $I_\mathcal{A}$ then becomes
\begin{align}
 &I_\mathcal{A} = \frac{\bar p_1\bar p_2}{2\pi} \int d\Omega_2 \frac{1}{\bar p_a} \Theta(\bar p_3+\bar p_4-\bar p_a) \Theta(\bar p_a-|\bar p_3-\bar p_4|)  \nonumber\\
 &\times\left(E_1E_2^*+\frac{\bar p_1^2+\bar p_2^2-\bar p_a^2}{2}\right) \left(E_3E_4^* +\frac{\bar p_3^2+\bar p_4^2-\bar p_a^2}{2}\right),
\end{align}
where the Heaviside functions come from the cosine limits \mbox{$-1\le x_3\le1$}. From Eq.~\eqref{eq:pa_x2} one can substitute the angular integral over $\Omega_2$ by
\begin{align}
 d\Omega_2=d\phi_2dx_2=d\phi_2d\bar p_a\frac{\bar p_a}{\bar p_1\bar p_2},
\end{align}
then the integral $I_\mathcal{A}$ takes the form
\begin{align}
\label{eq:ICstart}
 I_\mathcal{A} = & \int\limits_{p_{a-}}^{p_{a+}} d\bar p_a \left(E_1E_2^*+\frac{\bar p_1^2+\bar p_2^2-\bar p_a^2}{2}\right) \nonumber\\
 & \times\left(E_3E_4^* +\frac{\bar p_3^2+\bar p_4^2-\bar p_a^2}{2}\right).
\end{align}
The integration limits $p_{a-}$ and $p_{a+}$ arise from the combined constraints on $x_2$ and $x_3$,
\begin{align}
 p_{a-} = & \max\left\lbrace |\bar p_1-\bar p_2|,|\bar p_3-\bar p_4| \right\rbrace, \\
 p_{a+} = & \min\left\lbrace \bar p_1+\bar p_2,\bar p_3+\bar p_4 \right\rbrace.
\end{align}
Performing the integration over $\bar p_a$, $I_\mathcal{A}$ becomes 
\begin{align}
\label{eq:IA}
 I_\mathcal{A}=\frac{1}{60} & \left[3\left(p_{a+}^5-p_{a-}^5\right)-10\left(a+b\right)\left(p_{a+}^3-p_{a-}^3\right) \right.\nonumber\\
 &\quad \left.+60ab\left(p_{a+}-p_{a-}\right)\right],
\end{align}
where coefficients $a$ and $b$ are given by
\begin{align}
\label{eq:coeffA}
 a =E_1E_2^* +\frac{\bar p_1^2+\bar p_2^2}{2}, \;\; b =E_3E_4^* +\frac{\bar p_3^2+\bar p_4^2}{2}.
\end{align}

Taking similar steps, one can solve the angular integrals for all $I_\mathcal{X}$, and obtain 
\begin{subequations}
\begin{equation}
\begin{aligned}
\label{eq:IA}
 I_\mathcal{A}=\frac{1}{60}  [3\left(p_{a+}^5-p_{a-}^5\right)-10\left(a+b\right)\left(p_{a+}^3-p_{a-}^3\right)
+60ab(p_{a+}-p_{a-})],
\end{aligned}
\end{equation}
\begin{equation}
\label{eq:IB}
 I_\mathcal{B}=\frac{1}{60}  [3\left(p_{b+}^5-p_{b-}^5\right)-10\left(c+d\right)\left(p_{b+}^3-p_{b-}^3\right) 
+60cd(p_{b+}-p_{b-})],
\end{equation}
\begin{equation}
\begin{aligned}
 I_\mathcal{C} = & \Big[-\frac{\left(p_{a+}^7-p_{a-}^7\right)}{112}+\frac{a+\alpha_1}{20}\left(p_{a+}^5-p_{a-}^5\right)
 -\frac{a^2+4a\alpha_1-\alpha_0}{12}\left(p_{a+}^3-p_{a-}^3\right) \\
 & +\left(a^2\alpha_1-a\alpha_0\right)\left(p_{a+}-p_{a-}\right)-a^2\alpha_0\left(p_{a+}^{-1}-p_{a-}^{-1}\right)\Big], 
 \end{aligned}
\end{equation}
\begin{equation}
\begin{aligned}
\label{eq:SolID}
 I_\mathcal{D} = & \Big[\frac{\left(p_{c+}^7-p_{c-}^7\right)}{112}+\frac{e+\epsilon_1}{20}\left(p_{c+}^5-p_{c-}^5\right)
 +\frac{e^2+4e\epsilon_1+\epsilon_0}{12}\left(p_{c+}^3-p_{c-}^3\right) \\
 & +\left(e^2\epsilon_1+e\epsilon_0\right)\left(p_{c+}-p_{c-}\right)-e^2\epsilon_0\left(p_{c+}^{-1}-p_{c-}^{-1}\right)\Big],
\end{aligned}
\end{equation}
\begin{equation}
\begin{aligned}
\label{eq:IE}
 I_\mathcal{E}=\frac{1}{60} & \left[3\left(p_{a+}^5-p_{a-}^5\right)-20a\left(p_{a+}^3-p_{a-}^3\right) +60a^2\left(p_{a+}-p_{a-}\right)\right],
\end{aligned}
\end{equation}  
\begin{equation}
\label{eq:SolIF}
 I_\mathcal{F}=\frac{1}{60} [3\left(p_{c+}^5-p_{c-}^5\right)+20e\left(p_{c+}^3-p_{c-}^3\right)
 +60e^2\left(p_{c+}-p_{c-}\right)], 
\end{equation}
\begin{equation}
I_\mathcal{H} = \Big[\frac{p_{c+}^5-p_{c-}^5}{40} +\frac{e+2\epsilon_1}{12}\left(p_{c+}^3-p_{c-}^3\right)
 +\frac{2e\epsilon_1+\epsilon_0}{2}\left(p_{c+}-p_{c-}\right)-e\epsilon_0\left(p_{c+}^{-1}-p_{c-}^{-1}\right)\Big], 
\end{equation}
\begin{equation}
\label{eq:IJ}
 I_\mathcal{J}=\frac{1}{6} \left[-\left(p_{a+}^3-p_{a-}^3\right)+6a\left(p_{a+}-p_{a-}\right)\right],
\end{equation}
\begin{equation}
 I_\mathcal{K} = \frac{1}{6} \left[\left(p_{c+}^3-p_{c-}^3\right) +6e\left(p_{c+}-p_{c-}\right)\right],
\end{equation}
\begin{equation}
 I_\mathcal{L} = p_{a+}-p_{a-},
\end{equation}
\begin{equation}
I_\mathcal{L}^{AP} = \left\{ \begin{array}{cc} \frac{1}{\sqrt{Z_P}}\Big[\arctan\Big(\frac{p_{c+}}{\sqrt{Z_P}}\Big)-\arctan\Big(\frac{p_{c-}}{\sqrt{Z_P}}\Big)\Big],\;\;\; & \text{if~}Z_P>0, \\ \log(|\frac{1+\eta_P}{1-\eta_P}|)/(2\sqrt{|Z_P|}), \;\;\; & \text{if~}Z_P\le 0, \end{array} \right.
\end{equation}
\begin{equation}
I_\mathcal{K}^{AP} = \frac{1}{2} \Big[ (p_{c+}-p_{c-}) + (2 e-Z_P)I_\mathcal{L}^{AP} \Big],         
\end{equation}
\begin{equation}
\begin{aligned}
I_\mathcal{J}^{AP} =&  \epsilon_1 I_\mathcal{L}^{AP} + \frac{1}{4} \Big[ (p_{c+}-p_{c-}) -Z_P I_\mathcal{L}^{AP} \Big] - \frac{\epsilon_0}{Z_P} \Big( \frac{1}{p_{
c+}} - \frac{1}{p_{c-}} + I_\mathcal{L}^{AP} \Big)  \nonumber \\ 
 =& \Big( \epsilon_1 - \frac{\epsilon_0}{Z_P} - \frac{Z_P}{4}\Big) I_\mathcal{L}^{AP} + \Big[ \frac{1}{4}(p_{c+}-p_{c-}) - \frac{\epsilon_0}{Z_P} \Big( \frac{1}{p_{c+}} - \frac{1}{p_{c-}} \Big) \Big] ,  \\
\end{aligned}
\end{equation}       
\begin{equation}       
I_\mathcal{L}^{PP} = \frac{1}{2Z_P} \Big(  \frac{p_{c+}}{p_{c+}^2+Z_P} - \frac{p_{c-}}{p_{c-}^2+Z_P}  + I_\mathcal{L}^{AP} \Big), 
\end{equation}
\begin{align}
I_\mathcal{K}^{PP} = \Big(e-\frac{Z_P}{2}\Big) I_\mathcal{L}^{PP} + \frac{I_\mathcal{L}^{AP}}{2},  
\end{align}
\begin{equation}
\begin{aligned}
I_\mathcal{F}^{PP} =& e^2 I_\mathcal{L}^{PP} + e( -Z_P I_\mathcal{L}^{PP} + I_\mathcal{L}^{AP} ) + \frac{1}{4} \big[ Z_P^2 I_\mathcal{L}^{PP} - 2 Z_P I_\mathcal{L}^{AP} + (p_{c+}-p_{c-})
\big] \\
=& \Big(e-\frac{Z_P}{2}\Big)^2 I_\mathcal{L}^{PP}  + \Big(e-\frac{Z_P}{2}\Big) I_\mathcal{L}^{AP} + \frac{p_{c+}-p_{c-}}{4},   
\end{aligned}  
\end{equation}
\end{subequations}   
with 
\begin{equation}
\begin{aligned}
\label{eq:coeff_abcde}
& a =E_1E_2^* +\frac{\bar p_1^2+\bar p_2^2}{2}, \;\; b =E_3E_4^* +\frac{\bar p_3^2+\bar p_4^2}{2},  \\ 
& c = -E_1E_4^* +\frac{\bar p_1^2+\bar p_4^2}{2},\;\; d = -E_3E_2^* +\frac{\bar p_2^2+\bar p_3^2}{2},\;\;
e =  E_1E_3-\frac{\bar p_1^2+\bar p_3^2}{2} , \\
& \alpha_0 =  \frac{1}{4}\left(\bar p_1^2-\bar p_2^2\right)\left(\bar p_4^2-\bar p_3^2\right),\;\;
\alpha_1 =  E_1E_3-\frac{1}{4}\left(\bar p_1^2-\bar p_2^2+\bar p_3^2-\bar p_4^2\right), \\
& \epsilon_0 = \frac{1}{4}\left(\bar p_1^2-\bar p_3^2\right)\left(\bar p_2^2-\bar p_4^2\right),\;\;
\epsilon_1 =  E_1E_2^*+\frac{1}{4}\left(\bar p_1^2+\bar p_2^2-\bar p_3^2-\bar p_4^2\right), \\
& Z_P = m_\pi^2-m_3^2+2E_1E_3-\bar p_1^2-\bar p_3^2-2\Delta U(E_1-E_3)-\Delta U^2, \;\; \eta_P = \sqrt{|Z_P|} \frac{p_{c+}-p_{c-}}{p_{c+}p_{c-}+Z_P},  
\end{aligned}   
\end{equation} 
and  
\begin{equation}
\begin{aligned}
 p_{a-} = & \max\left\lbrace |\bar p_1-\bar p_2|,|\bar p_3-\bar p_4|\right\rbrace, \;\; p_{a+} =  \min\left\lbrace \bar p_1+\bar p_2,\bar p_3+\bar p_4 \right\rbrace \\
 p_{b-} = & \max\left\lbrace |\bar p_1-\bar p_4|,|\bar p_2-\bar p_3| \right\rbrace, \;\; p_{b+} = \min\left\lbrace \bar p_1+\bar p_4,\bar p_2+\bar p_3 \right\rbrace, \\
p_{c-} = & \max\left\lbrace |\bar p_1-\bar p_3|,|\bar p_2-\bar p_4| \right\rbrace, \;\;
p_{c+} =  \min\left\lbrace \bar p_1+\bar p_3,\bar p_2+\bar p_4 \right\rbrace.
\end{aligned}
\end{equation}  

The coefficients $\mathcal{X}, \mathcal{X}^{AP,PP}$ in Eq.~(\ref{eq:imfp-I}) are 
\begin{equation}
\begin{aligned}
 & \mathcal{A}=\left(g_V+g_A\right)^2+2g_AF_2\frac{m_2^*}{m_N}\Big(1-\frac{\Delta m^*}{2m_2^*}\Big), \;\; \mathcal{B}=\left(g_V-g_A\right)^2-2g_AF_2\frac{m_2^*}{m_N}\Big(1-\frac{\Delta m^*}{2m_2^*}\Big), \\
 & \mathcal{C}=\frac{F_2^2}{m_N^2}, \;\; \mathcal{D}=-\frac{F_2^2}{m_N^2}, \;\; \mathcal{E}=-\frac{F_2^2}{2m_N^2}[m_3^2-2\Delta U(E_3-E_1)+\Delta U^2], \\
 & \mathcal{F} = g_VF_2\frac{m_2^*}{m_N}\left(2-\frac{\Delta m_*}{m_2^*}\right) +\frac{F_2^2}{2m_N^2} \Big[ m_2^* m_4^* - Q_{24} + \frac{m_3^2}{4} - \Delta U(E_1 + E_2^*) - \frac{\Delta U^2}{4} \Big], \\
 & \mathcal{H} = \frac{F_2^2}{2m_N^2}\Big[ 2Q_{24} + m_3^2 + \Delta U(3E_1-E_3+2E_4^*) \Big], \;\; \mathcal{J} = g_VF_2\frac{\Delta m^*}{2m_N} \left[m_3^2 - \Delta U(E_1+E_3)\right] + \frac{F_2^2}{2 m_N^{2}} \mathcal{J}_{FF}, \\      
 &  \mathcal{K} = \left(g_A^2-g_V^2\right)m_2^*m_4^* + g_V F_2 \frac{m_2^*}{2m_N} \Big\{ -3m_3^2 + 4\Delta U(E_3-E_1)-\Delta U^2 + \frac{\Delta m^*}{m_2^*} [2Q_{24} + m_3^2 + \Delta U(2E_1-E_3+E_4^*)] \Big\}  \\
 &\qquad+ \frac{F_2^2}{2m_N^2} \mathcal{K}_{FF}, \\   
 & \mathcal{L} =  g_vF_2\frac{m_2^*}{m_N} \Delta U E_1 \Big[ m_3^2 -\Delta U E_3 + \frac{\Delta m^*}{2m_2^*} \Big(-Q_{24} -\frac{m_3^2}{2}-\Delta U E_4^* +\frac{\Delta U^2}{2}\Big)\Big] + \frac{F_2^2}{2m_N^2}\mathcal{L}_{FF},  \\
& \mathcal{F}^{PP} = 2m_N^2 g_A^2 (m_3^2 - \Delta U^2), \;\; \mathcal{J}^{AP} = 2m_N g_A^2 \Delta m^*[ m_3^2-\Delta U(E_1+E_3) ],  \\
& \mathcal{K}^{AP} = 2 g_A^2 m_N \Big[ m_2^*(\Delta U^2-m_3^2) + \Delta m^* \Delta U(E_1 + E_2^*)\Big],  \\
& \mathcal{K}^{PP} = 2 g_A^2 m_N^2 \Big[ \frac{m_3^2}{2}(-m_3^2+\Delta m^{*2}) + \Delta U m_3^2 (E_3-3E_1)+\Delta U^2\Big( 2E_1E_3-\frac{\Delta m^{*2}}{2}\Big) + \Delta U^3(E_4^*-E_2^*) - \frac{\Delta U^4}{2} \Big], \\  
& \mathcal{L}^{AP} = 2 g_A^2 m_N \Delta U E_1\Big[  2 m_2^*m_3^2 - \Delta m^*\Big(Q_{24}+\frac{m_3^2}{2}\Big) - \Delta U\big(2E_3 m_2^*+E_4^* \Delta m^*\big)+\frac{\Delta m^*}{2}\Delta U^2\Big],  \\
&  \mathcal{L}^{PP} =  2 g_A^2 m_N^2 \Delta U E_1\Big\{  m_3^2(m_3^2 -\Delta m^{*2}) + \Delta U\big[  -m_3^2(3E_3-2E_1) + \Delta m^{*2} E_3 \big] + \Delta U^2[m_3^2 + 2E_3(E_3-E_1)]-E_3\Delta U^3 \Big\},  
\end{aligned}
\end{equation} 
where 
\begin{equation}
\begin{aligned}
\mathcal{J}_{FF} =& \Delta U \Big\{ -m_3^2\Big( E_1 + \frac{E_2^*+E_4^*}{2}\Big) + Q_{24}(E_3-3E_1) + \frac{\Delta U}{2} \big[ E_4^*(3E_3-5E_1) + E_2^*(E_3+E_1) + E_3^2-E_1^2 - 2Q_{24}\big] \\
& + \Delta U^2\Big( E_1 - E_3 - E_4^*\Big) + \frac{\Delta U^3}{2} \Big\}, \\
\mathcal{K}_{FF} =& -(m^*_2+3m^*_4)m^*_2 \frac{m_3^2}{4} + Q_{24}^2 + Q_{24}\frac{m_3^2}{4}-\frac{m_3^4}{8} \\ 
&+ \Delta U\Big[ \frac{Q_{24}}{2}(3E_1-E_2^*+E_3+3E_4^*)+\frac{m_3^2}{4}( 2E_2^* + E_3 + E_1) + m^*_2m^*_4(E_3-E_1) \Big]  \\
&+\Delta U^2 \Big[  \frac{1}{4}( m_2^{*2}-m^*_2m^*_4-3Q_{24})  + \frac{E_4^*}{2} (E_3+2E_1-E_2^*+E_4^*) + E_2^*\Big(\frac{1}{2}E_1-E_3\Big) + \frac{E_1^2}{2}\Big] \\
& + \frac{\Delta U^3}{4} ( -E_1 + 2E_2^* - E_3 - 2E_4^*) + \frac{\Delta U^4}{8},  \\
\mathcal{L}_{FF} = & \frac{\Delta U E_1}{4} \Big\{ m_3^2(m^*_2 + m^*_4)^2 - 4Q_{24}^2 + \Delta U[ -m_3^2( E_2^*+E_4^*+E_1) - 2E_3(m_2^{*2}+m^*_2m^*_4) + 2Q_{24}(E_2^*-3E_4^*-E_1) ]  \\
& + 2\Delta U^2[  E_2^*E_3 + E_4^*( E_2^*-E_4^*-E_1) + Q_{24}] + \Delta U^3 [  -E_2^* + E_4^* + E_1 ] \Big\},
\end{aligned}
\end{equation}
with $\Delta U=U_2-U_4$, $\Delta m^* = m_2^*-m_4^*$, and $Q_{24} = (m_2^{*2}-m_4^{*2})/2$.

\section{4D integrals and physically allowed region}   
\label{sec:append_bounds}

The 2D integrals in this work are done straightforwardly via the
Gauss-Legendre quadrature in each dimension. In this subsection, we
will show the details of computing the opacities from the 4D
integrals. We need to determine the boundaries of the
integration variables that are kinematically allowed by energy-momentum
conservation.

\subsection{bounds for neutrino absorption or scattering}  
\label{sec:bound_cap} 

The discussions below can apply equally to both neutrino scattering and absorption on nucleons and leptons. We take neutrino absorption on nucleons, $\nu_1+N_2\to l_3+N_4$,
for demonstration. It can be easily seen, for example, by taking $N_{2,4} \to l_{2,4}$ and $l_3\to \nu_3$, the results for neutrino scattering on leptons are obtained.
 
For neutrino absorption on nucleons, we set $\bm{q}= \bm{p}_4-\bm{p_2}=\bm{p}_1-\bm{p}_3$ along the $z$-axis, and choose $E_2$, $|\bm{q}|$, $\cos\theta_{q2}$ and $\phi_{q2}$ as the integration variables, where $\theta_{q2}$ and $\phi_{q2}$ are the pole and azimuthal angles of $\bm{p}_2$ with respect to $\bm{q}$ or the $z$-axis. Once the four variables are fixed, the four-momenta of all the particles can be determined as follows: $\bm{p}_2$ is fixed when $E_2$, $\theta_{q2}$ and $\phi_{q2}$ are known; furthermore, $\bm{p}_4$ and then $q^0=E_4-E_2$ can be determined if $\bm{q}=|\bm{q}| \bm{\hat z}$ is known. For a given neutrino energy $E_1$ with $q^0$ and $\bm{q}$ fixed, $\bm{p}_{1, 3}$ are then uniquely determined.   

Since $q_0 = \sqrt{ |\bm{p}_2+\bm{q}|^2 + m_4^{*2} } + U_4  - E_2 = E_1 - \sqrt{ |\bm{p}_1-\bm{q}|^2 + m_3^{2} }$,
the maximal and minimal values of $q_0$ can be expressed as    
\begin{subequations} 
\label{eq:q0_minmax}
\begin{align}
& q_0^{\rm H, max} = \sqrt{ (|\bm{p}_2|+|\bm{q}|)^2 + m_4^{*2} } + U_4  - E_2,  \\   
& q_0^{\rm H, min} = \sqrt{ (|\bm{p}_2|-|\bm{q}|)^2 + m_4^{*2} } + U_4  - E_2,    \\
& q_0^{\rm L, max} = E_1 - \sqrt{ (E_1-|\bm{q}|)^2 + m_3^{2} },   \\
& q_0^{\rm L, min} = E_1 - \sqrt{ (E_1+|\bm{q}|)^2 + m_3^{2} }.    
\end{align}
\end{subequations}
For the kinematically allowed values of $|\bm{p}_2|$ and $|\bm{q}|$, we should have $q_0^{\rm L, max} \ge q_0^{\rm H, min}$ and meanwhile, $q_0^{\rm H, max} \ge q_0^{\rm L, min}$. From these requirements, we can obtain the upper and lower bounds of the integration variables.

\noindent
(a) Allowed range of $E_1$ and upper/lower bounds of $E_2$ or $|\bm{p}_2|$:

We note that $q_0^{\rm H, max}$ grows with $|\bm{q}|$ while
$q_0^{\rm L, min}$ decreases, so we can always find the allowed region
of $|\bm{q}|$ for any given value of $|\bm{p}_2|$ that satisfy the
inequality $q_0^{\rm H, max} \ge q_0^{\rm L, min}$. In other words, the allowed range of
$|\bm{p}_2|$ can be determined by requiring that there are solutions
of $|\bm{q}|$ that satisfy
$F(|\bm{q}|, |\bm{p}_2|)=q_0^{\rm H, min}-q_0^{\rm L, max}\le 0$. From
\begin{align}
\frac{\partial F(|\bm{q}|, |\bm{p}_2|)}{\partial |\bm{q}|} =0,
\end{align}   
one can find that $F(|\bm{q}|, |\bm{p}_2|)$ reaches its minimum at 
\begin{equation}
|\bm{q}|_{\rm min}=\frac{m_3 |\bm{p}_2|+m_4^* E_1}{\widetilde m_4^*},
\end{equation} 
with $\widetilde m_4^* = m_4^*+m_3$. For $\widetilde m_4^* < m_2^*$, $F(|\bm{q}|_{\rm min}, |\bm{p}_2|)$ first decreases with $|\bm{p}_2|$, reaches its minimum at 
\begin{equation}
|\bm{p}_2|_{\rm min}=\frac{E_1m_2^*}{m_2^*-\widetilde m_4^*},
\end{equation} 
and then increases. For $\widetilde m_4^* \ge m_2^*$, $F(|\bm{q}|_{\rm min}, |\bm{p}_2|)$ decreases monotonically. 

Since the minimum of $F(|\bm{q}|, |\bm{p}_2|)$ should be negative, the allowed values of $E_1$ for a nonzero reaction rate should satisfy    
\begin{equation}
\begin{aligned}
\label{eq:cond_E1a}
F(|\bm{q}|_{\rm min}, |\bm{p}_2|_{\rm min}) 
=\sqrt{(|\bm{p}_2|_{\rm min}-E_1)^2+\widetilde m_4^{*2}} - 
\sqrt{|\bm{p}_2|^2_{\rm min}+m_2^{*2}} 
+U_4-U_2-E_1< 0,
\end{aligned}    
\end{equation} 
if $\widetilde m_4^* < m_2^*$, or  
\begin{equation}
\begin{aligned}
\label{eq:cond_E1b}
F(|\bm{q}|_{\rm min}, \infty) = U_4-U_2-2E_1< 0,
\end{aligned}
\end{equation}     
if $\widetilde m_4^* \ge m_2^*$.

If the maximum of $F(|\bm{q}|_{\rm min}, |\bm{p}_2|)$ is negative, i.e., 
\begin{align}
\max\{F(|\bm{q}|_{\rm min}, 0), F(|\bm{q}|_{\rm min}, \infty)\}\le 0, 
\end{align}
there will be no constraint on $|\bm{p}_2|$. Otherwise, the
bounds are given by the positive solutions of $F(|\bm{q}|_{\rm min}, |\bm{p}_2|)=0$.
We have checked that the lower bound of $|\bm{p}_2|$ is given by 
\begin{equation}
|\bm{p}_2|^{\rm low} = \frac{E_1 E_{sq} \pm \sqrt{ E'^2 E_{qu}}}{2(E_1^2 - E'^2)}, \label{eq:p2_bound} 
\end{equation}  
where `$+$' and `$-$' correspond to cases with $U_4-U_2-E_1\ge 0$ and $U_4-U_2-E_1<0$, respectively, and
\begin{equation}
\begin{aligned}
& E' = E_1+U_2-U_4, \\ 
& E_{sq} = E_1^2+\widetilde m_4^{*2}-m_2^{*2}-E'^2, \\
& E_{qu} = E_{sq}^2+4m_2^{*2}(E_1^2 - E'^2).
\end{aligned}
\end{equation}
In the special case where $E_1^2 = E'^2$, the possible solution is  
\begin{equation}
|\bm{p}_2|^{\rm low} = \frac{E_{sq}^2-4 E'^2m_2^{*2}}{4 E_1E_{sq}}. \label{eq:p2_0}  
\end{equation}

\noindent
(b) Upper/lower bounds of $|\bm{q}|$: 

For a given value of $|\bm{p}_2|$ between $|\bm{p}_2|^{\rm low}$ and $|\bm{p}_2|^{\rm up}$, the bounds of
$|\bm{q}|$ can be determined by equations $q_0^{\rm H,min}=q_0^{\rm L,max}$ or $q_0^{\rm H,max}=q_0^{\rm L,min}$, see Eq.~(\ref{eq:q0_minmax}), and we obtain
\begin{equation}
\begin{aligned}
\label{eq:q_bound}
& |\bm{q}|^{\rm low} = \Bigg|\frac{B_q + \sqrt{B_q^2 - 4 A_q C_q}}{2 A_q}\Bigg|, \\
& |\bm{q}|^{\rm up} = \frac{-B_q + \sqrt{B_q^2 - 4 A_q C_q}}{2 A_q},   
\end{aligned}    
\end{equation}
where
\begin{equation}
\label{eq:ABC_q}
\begin{aligned}
&A_q =  4[ E_q^2 - (E_1-|\bm{p}_2|)^2 ], \\
&B_q = 4[ E_1(E_{qsq}-E_q^2)-|\bm{p}_2|(E_{qsq}+E_q^2) ], \\ 
&C_q = 4E_q^2( |\bm{p}_2|^2+m_4^{*2} ) - (  E_{qsq}-E_q^2 )^2,   
\end{aligned}
\end{equation}
with $E_q=E_1+E_2-U_4$ and $E_{qsq}=E_1^2+m_3^2-m_4^{*2}-|\bm{p}_2|^2$.

\noindent
(c) Upper/lower bounds of $\cos\theta_{q2}$: 

Once fixing the values of $\bm{p}_2$ and $|\bm{q}|$, the minimum and maximum of $q_0$
are simply 
\begin{equation}
\begin{aligned}
q_0^{\rm min}=\max\{q_0^{\rm H, min}, q_0^{\rm L, min}\}, \\
q_0^{\rm max}=\min\{q_0^{\rm H, max}, q_0^{\rm L, max}\}.
\end{aligned} 
\end{equation} 
Since $q_0=E_4 - E_2 = \sqrt{ |\bm{p}_2+\bm{q}|^2 + m_4^{*2}} + U_4 - E_2$, the bounds of $\cos\theta_{q2}$ can be obtained as
\begin{equation}
\begin{aligned}
\label{eq:costheta_bound}
& \cos\theta_{q2}^{\rm low}=\max\Bigg\{-1, 
 \frac{ (q_0^{\rm min}+E_2-U_4)^2-m_4^{*2}-|\bm{p}_2|^2-|\bm{q}|^2}{2 |\bm{p}_2||\bm{q}|}\Bigg\}, \\
& \cos\theta_{q2}^{\rm up}=\min\Bigg\{1, 
 \frac{ (q_0^{\rm max}+E_2-U_4)^2-m_4^{*2}-|\bm{p}_2|^2-|\bm{q}|^2}{2 |\bm{p}_2||\bm{q}|}\Bigg\}.
\end{aligned} 
\end{equation} 

\noindent
(d) Upper/lower bounds of $\phi_{q2}$:
\begin{align}
\label{eq:phi_bound}
\phi_{q2}^{\rm low}= 0, \;\; \phi_{q2}^{\rm up} = 2\pi.  
\end{align} 

In summary, Eqs.~(\ref{eq:cond_E1a}) or (\ref{eq:cond_E1b}) determines the threshold of neutrino energy $E_1$, and Eqs.~(\ref{eq:p2_bound}) and \eqref{eq:p2_0}, (\ref{eq:q_bound}), (\ref{eq:costheta_bound}), and (\ref{eq:phi_bound}) give the upper/lower bounds for the integration variables $|E_2|$, $|\bm{q}|$, $\cos\theta_{q2}$ and $\phi_{q2}$, respectively.     
 
\subsection{bounds for inverse decay}  
\label{sec:bound_decay}

The bounds of the integration variables for inverse decay, $\nu_1 + N_2 + l_3 \to N_4$, or $\nu_1 + l_2 + \nu_3 \to l_4$, can be obtained similarly as for neutrino absorption or scattering, as discussed above. For inverse decay, we have different definitions of $\bm{q}$ and $q_0$ with $\bm{q}= \bm{p}_4-\bm{p}_2=\bm{p}_1+\bm{p}_3$ and $q_0= E_4-E_2=E_1+E_3$. Correspondingly, the maximal and minimal values of $q_0$ shown in Eq.~(\ref{eq:q0_minmax}) turn to
\begin{subequations} 
\label{eq:q0_minmax_dec}
\begin{align}
& q_{0, \rm dec}^{\rm H, max} = \sqrt{ (|\bm{p}_2|+|\bm{q}|)^2 + m_4^{*2} } + U_4  - E_2,  \\   
& q_{0, \rm dec}^{\rm H, min} = \sqrt{ (|\bm{p}_2|-|\bm{q}|)^2 + m_4^{*2} } + U_4  - E_2,    \\
& q_{0, \rm dec}^{\rm L, max} = E_1 + \sqrt{ E_1+|\bm{q}|)^2 + m_3^{2} },   \\
& q_{0, \rm dec}^{\rm L, min} = E_1 + \sqrt{ (E_1-|\bm{q}|)^2 + m_3^{2} }.    
\end{align}
\end{subequations} 
We note that $q_{0, \rm dec}^{\rm L, max} > q_{0, \rm dec}^{\rm H, min}$ is guaranteed at $|\bm{q}| \to \infty$, so we only need to consider the requirement $q_{0, \rm dec}^{\rm H, max} \ge q_{0, \rm dec}^{\rm L, min}$. We choose the same integration variables as for neutrino scattering/absorption and take the same steps to determine their bounds.
 
\noindent
(a) Allowed range of $E_1$ and upper/lower bounds of $E_2$ or $|\bm{p}_2|$:

If $\widetilde m_{4, \rm dec}^*= m_4^*-m_3>m_2^*$, the kinematically allowed $E_1$ should satisfy 
\begin{equation}
\begin{aligned}
F_{\rm dec}(|\bm{q}|_{\rm max}, |\bm{p}_2|_{\rm max}) 
=\sqrt{(|\bm{p}_2|_{\rm max}+E_1)^2+\widetilde m_{4, \rm dec}^{*2}} - 
\sqrt{|\bm{p}_2|^2_{\rm max}+m_2^{*2}} 
+U_4-U_2-E_1> 0,
\end{aligned}    
\end{equation}   
where $F_{\rm dec}( |\bm{q}|, |\bm{p}_2|) = q_{0, \rm dec}^{\rm H, max}-q_{0, \rm dec}^{\rm L, min}$, and 
\begin{align}
& |\bm{q}|_{\rm max} =  \frac{m_3|\bm{p}_2| + m_4^* E_1}{\widetilde m_{4,\rm dec}^*}, \\
& |\bm{p}_2|_{\rm max} = \frac{m_2^* E_1}{\widetilde m_{4, \rm dec}^* - m_2^*}.
\end{align}
While for $\widetilde m_{4, \rm dec}^* \le m_2^*$, one should have  
\begin{align}
F_{\rm dec}( |\bm{q}|_{\rm max}, \infty) = U_4 - U_2 > 0.  
\end{align}

If the minimum of $F_{\rm dec}(|\bm{q}|_{\rm max}, |\bm{p}_2|)$ is positive, i.e., 
\begin{align}
\min\{F_{\rm dec}(|\bm{q}|_{\rm max}, 0), F_{\rm dec}(|\bm{q}|_{\rm max}, \infty)\} \ge 0, 
\end{align}  
there will be no constraint on $|\bm{p}_2|$. Otherwise, the lower
bound of $|\bm{p}_2|$ is given by the positive solution of $F_{\rm dec}(|\bm{q}|, |\bm{p}_2|)=0$: 
\begin{equation}
|\bm{p}_2|^{\rm low} = \frac{E_1 E_{sq, \rm dec} \pm \sqrt{ E'^2_{\rm dec} E_{qu, \rm dec}}}{2(E'^2_{\rm dec} - E_1^2)}, \label{eq:p2_pm_dec} 
\end{equation}  
where `$+$' and `$-$' correspond to cases with $U_4-U_2-E_1\ge 0$ and $U_4-U_2-E_1<0$, respectively, and  
\begin{equation}
\begin{aligned}
& E'_{\rm dec} = E_1+U_2-U_4, \\ 
& E_{sq, \rm dec} = E_1^2+\widetilde m_{4, \rm dec}^{*2}-m_2^{*2}-E'^2_{\rm dec}, \\
& E_{qu, \rm dec} = E_{sq, \rm dec}^2+4m_2^{*2}(E_{1}^2 - E'^2_{\rm dec}).
\end{aligned}
\end{equation}
In the case where $E_1^2 = E'^2_{\rm dec}$, 
\begin{equation}
|\bm{p}_2|^{\rm low} =\frac{4 E'^2m_2^{*2}-E_{sq}^2}{4 E_1E_{sq}}. \label{eq:p2_0_dec}  
\end{equation}
 
\noindent
(b) Upper/lower bounds of $|\bm{q}|$:
\begin{equation}
\begin{aligned}
\label{eq:q_bound_d}
& |\bm{q}|^{\rm low} = \Bigg|\frac{B_{q, \rm dec} + \sqrt{B_{q, \rm dec}^2 - 4 A_{q, \rm dec} C_{q, \rm dec}}}{2 A_{q, \rm dec}}\Bigg|, \\
& \widetilde{|\bm{q}|}^{\rm up} = \frac{-B_{q, \rm dec} + \sqrt{B_{q, \rm dec}^2 - 4 A_{q, \rm dec} C_{q, \rm dec}}}{2 A_{q, \rm dec}},   
\end{aligned}    
\end{equation}
where
\begin{equation}
\begin{aligned}
&A_{q, \rm dec} =  4[ E_q^2 - (E_1+|\bm{p}_2|)^2 ], \\
&B_{q, \rm dec} = 4[ E_1(E_{qsq}-E_q^2)+|\bm{p}_2|(E_{qsq}+E_q^2) ], \\ 
&C_{q, \rm dec} = 4E_q^2( |\bm{p}_2|^2+m_4^{*2} ) - (  E_{qsq}-E_q^2 )^2,   
\end{aligned}
\end{equation}
with $E_q$ and $E_{qsq}$ given below Eq.~(\ref{eq:ABC_q}). It should emphasized that
the upper bound $|\bm{q}|^{\rm up}=\widetilde{|\bm{q}|}^{\rm up}$ only if $\widetilde{|\bm{q}|}^{\rm up}>0$.
Otherwise, there will be no upper bound for $|\bm{q}|$. 
 
\noindent
(c) Upper/lower bounds of $\cos\theta_{q2}$: 
\begin{equation}
\begin{aligned}
\label{eq:costheta_bound_d}
& \cos\theta_{q2}^{\rm low}=\max\Bigg\{-1, 
 \frac{ (q_{0, \rm dec}^{\rm min}+E_2-U_4)^2-m_4^{*2}-|\bm{p}_2|^2-|\bm{q}|^2}{2 |\bm{p}_2||\bm{q}|}\Bigg\}, \\
& \cos\theta_{q2}^{\rm up}=\min\Bigg\{1, 
 \frac{ (q_{0, \rm dec}^{\rm max}+E_2-U_4)^2-m_4^{*2}-|\bm{p}_2|^2-|\bm{q}|^2}{2 |\bm{p}_2||\bm{q}|}\Bigg\},
\end{aligned} 
\end{equation}  
where $q_{0, \rm dec}^{\rm min}=\max\{q_{0, \rm dec}^{\rm H, min}, q_{0, \rm dec}^{\rm L, min}\}$ and
$q_{0, \rm dec}^{\rm max}=\min\{q_{0, \rm dec}^{\rm H, max}, q_{0, \rm dec}^{\rm L, max}\}$, with $q_{0, \rm dec}^{\rm H/L, max/min}$ shown in Eq.~(\ref{eq:q0_minmax_dec}).

\noindent
(d) Upper/lower bounds of $\phi_{q2}$:
\begin{align}
\label{eq:phi_bound_d}
\phi_{q2}^{\rm low}= 0, \;\; \phi_{q2}^{\rm up} = 2\pi.  
\end{align} 

\end{widetext}


\begin{thebibliography}{61}%
\makeatletter
\providecommand \@ifxundefined [1]{%
 \@ifx{#1\undefined}
}%
\providecommand \@ifnum [1]{%
 \ifnum #1\expandafter \@firstoftwo
 \else \expandafter \@secondoftwo
 \fi
}%
\providecommand \@ifx [1]{%
 \ifx #1\expandafter \@firstoftwo
 \else \expandafter \@secondoftwo
 \fi
}%
\providecommand \natexlab [1]{#1}%
\providecommand \enquote  [1]{``#1''}%
\providecommand \bibnamefont  [1]{#1}%
\providecommand \bibfnamefont [1]{#1}%
\providecommand \citenamefont [1]{#1}%
\providecommand \href@noop [0]{\@secondoftwo}%
\providecommand \href [0]{\begingroup \@sanitize@url \@href}%
\providecommand \@href[1]{\@@startlink{#1}\@@href}%
\providecommand \@@href[1]{\endgroup#1\@@endlink}%
\providecommand \@sanitize@url [0]{\catcode `\\12\catcode `\$12\catcode
  `\&12\catcode `\#12\catcode `\^12\catcode `\_12\catcode `\%12\relax}%
\providecommand \@@startlink[1]{}%
\providecommand \@@endlink[0]{}%
\providecommand \url  [0]{\begingroup\@sanitize@url \@url }%
\providecommand \@url [1]{\endgroup\@href {#1}{\urlprefix }}%
\providecommand \urlprefix  [0]{URL }%
\providecommand \Eprint [0]{\href }%
\providecommand \doibase [0]{http://dx.doi.org/}%
\providecommand \selectlanguage [0]{\@gobble}%
\providecommand \bibinfo  [0]{\@secondoftwo}%
\providecommand \bibfield  [0]{\@secondoftwo}%
\providecommand \translation [1]{[#1]}%
\providecommand \BibitemOpen [0]{}%
\providecommand \bibitemStop [0]{}%
\providecommand \bibitemNoStop [0]{.\EOS\space}%
\providecommand \EOS [0]{\spacefactor3000\relax}%
\providecommand \BibitemShut  [1]{\csname bibitem#1\endcsname}%
\let\auto@bib@innerbib\@empty
\bibitem [{\citenamefont {{Janka}}(2012)}]{Janka:2012}%
  \BibitemOpen
  \bibfield  {author} {\bibinfo {author} {\bibfnamefont {H.-T.}\ \bibnamefont
  {{Janka}}},\ }\href {\doibase 10.1146/annurev-nucl-102711-094901} {\bibfield
  {journal} {\bibinfo  {journal} {Annu. Rev. Nucl. Part. Sci.}\ }\textbf
  {\bibinfo {volume} {62}},\ \bibinfo {pages} {407} (\bibinfo {year}
  {2012})}\BibitemShut {NoStop}%
\bibitem [{\citenamefont {{Burrows}}\ \emph {et~al.}(2006)\citenamefont
  {{Burrows}}, \citenamefont {{Reddy}},\ and\ \citenamefont
  {{Thompson}}}]{Burrows.Reddy.Thompson:2006}%
  \BibitemOpen
  \bibfield  {author} {\bibinfo {author} {\bibfnamefont {A.}~\bibnamefont
  {{Burrows}}}, \bibinfo {author} {\bibfnamefont {S.}~\bibnamefont {{Reddy}}},
  \ and\ \bibinfo {author} {\bibfnamefont {T.~A.}\ \bibnamefont {{Thompson}}},\
  }\href {\doibase 10.1016/j.nuclphysa.2004.06.012} {\bibfield  {journal}
  {\bibinfo  {journal} {Nucl. Phys.}\ ,\ \bibinfo {pages} {356}} (\bibinfo
  {year} {2006})}\BibitemShut {NoStop}%
\bibitem [{\citenamefont {Janka}\ \emph {et~al.}(2007)\citenamefont {Janka},
  \citenamefont {Langanke}, \citenamefont {Marek}, \citenamefont
  {Mart{\'i}nez-Pinedo},\ and\ \citenamefont
  {M{\"u}ller}}]{Janka.Langanke.ea:2007}%
  \BibitemOpen
  \bibfield  {author} {\bibinfo {author} {\bibfnamefont {H.-T.}\ \bibnamefont
  {Janka}}, \bibinfo {author} {\bibfnamefont {K.}~\bibnamefont {Langanke}},
  \bibinfo {author} {\bibfnamefont {A.}~\bibnamefont {Marek}}, \bibinfo
  {author} {\bibfnamefont {G.}~\bibnamefont {Mart{\'i}nez-Pinedo}}, \ and\
  \bibinfo {author} {\bibfnamefont {B.}~\bibnamefont {M{\"u}ller}},\ }\href
  {\doibase 10.1016/j.physrep.2007.02.002} {\bibfield  {journal} {\bibinfo
  {journal} {Phys. Rep.}\ }\textbf {\bibinfo {volume} {442}},\ \bibinfo {pages}
  {38} (\bibinfo {year} {2007})}\BibitemShut {NoStop}%
\bibitem [{\citenamefont {Burrows}(2013)}]{Burrows:2013}%
  \BibitemOpen
  \bibfield  {author} {\bibinfo {author} {\bibfnamefont {A.}~\bibnamefont
  {Burrows}},\ }\href {\doibase 10.1103/RevModPhys.85.245} {\bibfield
  {journal} {\bibinfo  {journal} {Rev. Mod. Phys.}\ }\textbf {\bibinfo {volume}
  {85}},\ \bibinfo {pages} {245} (\bibinfo {year} {2013})}\BibitemShut
  {NoStop}%
\bibitem [{\citenamefont {Mart{\'{i}}nez-Pinedo}\ \emph
  {et~al.}(2016)\citenamefont {Mart{\'{i}}nez-Pinedo}, \citenamefont {Fischer},
  \citenamefont {Langanke}, \citenamefont {Lohs}, \citenamefont {Sieverding},\
  and\ \citenamefont {Wu}}]{Martinez-Pinedo.Fischer.ea:2016}%
  \BibitemOpen
  \bibfield  {author} {\bibinfo {author} {\bibfnamefont {G.}~\bibnamefont
  {Mart{\'{i}}nez-Pinedo}}, \bibinfo {author} {\bibfnamefont {T.}~\bibnamefont
  {Fischer}}, \bibinfo {author} {\bibfnamefont {K.}~\bibnamefont {Langanke}},
  \bibinfo {author} {\bibfnamefont {A.}~\bibnamefont {Lohs}}, \bibinfo {author}
  {\bibfnamefont {A.}~\bibnamefont {Sieverding}}, \ and\ \bibinfo {author}
  {\bibfnamefont {M.-R.}\ \bibnamefont {Wu}},\ }\enquote {\bibinfo {title}
  {{Neutrinos and Their Impact on Core-Collapse Supernova Nucleosynthesis}},}\
  in\ \href {\doibase 10.1007/978-3-319-20794-0_78-1} {\emph {\bibinfo
  {booktitle} {Handbook of Supernovae}}},\ \bibinfo {editor} {edited by\
  \bibinfo {editor} {\bibfnamefont {A.~W.}\ \bibnamefont {{Alsabti}}}\ and\
  \bibinfo {editor} {\bibfnamefont {P.}~\bibnamefont {{Murdin}}}}\ (\bibinfo
  {publisher} {Springer International Publishing},\ \bibinfo {address} {Cham},\
  \bibinfo {year} {2016}),\ pp.\ \bibinfo {pages} {1805--1841}\BibitemShut
  {NoStop}%
\bibitem [{\citenamefont {Melson}\ \emph {et~al.}(2015)\citenamefont {Melson},
  \citenamefont {Janka}, \citenamefont {Bollig}, \citenamefont {Hanke},
  \citenamefont {Marek},\ and\ \citenamefont
  {M{\"{u}}ller}}]{Melson.Janka.ea:2015}%
  \BibitemOpen
  \bibfield  {author} {\bibinfo {author} {\bibfnamefont {T.}~\bibnamefont
  {Melson}}, \bibinfo {author} {\bibfnamefont {H.-T.}\ \bibnamefont {Janka}},
  \bibinfo {author} {\bibfnamefont {R.}~\bibnamefont {Bollig}}, \bibinfo
  {author} {\bibfnamefont {F.}~\bibnamefont {Hanke}}, \bibinfo {author}
  {\bibfnamefont {A.}~\bibnamefont {Marek}}, \ and\ \bibinfo {author}
  {\bibfnamefont {B.}~\bibnamefont {M{\"{u}}ller}},\ }\href {\doibase
  10.1088/2041-8205/808/2/L42} {\bibfield  {journal} {\bibinfo  {journal}
  {Astrophys. J.}\ }\textbf {\bibinfo {volume} {808}},\ \bibinfo {pages} {L42}
  (\bibinfo {year} {2015})}\BibitemShut {NoStop}%
\bibitem [{\citenamefont {{Schinder}}\ and\ \citenamefont
  {{Shapiro}}(1982)}]{Schinder.Shapiro:1982}%
  \BibitemOpen
  \bibfield  {author} {\bibinfo {author} {\bibfnamefont {P.~J.}\ \bibnamefont
  {{Schinder}}}\ and\ \bibinfo {author} {\bibfnamefont {S.~L.}\ \bibnamefont
  {{Shapiro}}},\ }\href {\doibase 10.1086/190818} {\bibfield  {journal}
  {\bibinfo  {journal} {Astrophys. J. Suppl.}\ }\textbf {\bibinfo {volume}
  {50}},\ \bibinfo {pages} {23} (\bibinfo {year} {1982})}\BibitemShut {NoStop}%
\bibitem [{\citenamefont {{Bruenn}}(1985)}]{Bruenn:1985}%
  \BibitemOpen
  \bibfield  {author} {\bibinfo {author} {\bibfnamefont {S.~W.}\ \bibnamefont
  {{Bruenn}}},\ }\href {\doibase 10.1086/191056} {\bibfield  {journal}
  {\bibinfo  {journal} {Astrophys. J. Suppl.}\ }\textbf {\bibinfo {volume}
  {58}},\ \bibinfo {pages} {771} (\bibinfo {year} {1985})}\BibitemShut
  {NoStop}%
\bibitem [{\citenamefont {{Mezzacappa}}\ and\ \citenamefont
  {{Bruenn}}(1993)}]{Mezzacappa.Bruenn:1993}%
  \BibitemOpen
  \bibfield  {author} {\bibinfo {author} {\bibfnamefont {A.}~\bibnamefont
  {{Mezzacappa}}}\ and\ \bibinfo {author} {\bibfnamefont {S.~W.}\ \bibnamefont
  {{Bruenn}}},\ }\href {\doibase 10.1086/172791} {\bibfield  {journal}
  {\bibinfo  {journal} {\apj}\ }\textbf {\bibinfo {volume} {410}},\ \bibinfo
  {pages} {740} (\bibinfo {year} {1993})}\BibitemShut {NoStop}%
\bibitem [{\citenamefont {Hannestad}\ and\ \citenamefont
  {Raffelt}(1998)}]{Hannestad.Raffelt:1998}%
  \BibitemOpen
  \bibfield  {author} {\bibinfo {author} {\bibfnamefont {S.}~\bibnamefont
  {Hannestad}}\ and\ \bibinfo {author} {\bibfnamefont {G.}~\bibnamefont
  {Raffelt}},\ }\href {\doibase 10.1086/306303} {\bibfield  {journal} {\bibinfo
   {journal} {Astrophys. J.}\ }\textbf {\bibinfo {volume} {507}},\ \bibinfo
  {pages} {339} (\bibinfo {year} {1998})}\BibitemShut {NoStop}%
\bibitem [{\citenamefont {Reddy}\ \emph {et~al.}(1998)\citenamefont {Reddy},
  \citenamefont {Prakash},\ and\ \citenamefont
  {Lattimer}}]{Reddy.Prakash.Lattimer:1998}%
  \BibitemOpen
  \bibfield  {author} {\bibinfo {author} {\bibfnamefont {S.}~\bibnamefont
  {Reddy}}, \bibinfo {author} {\bibfnamefont {M.}~\bibnamefont {Prakash}}, \
  and\ \bibinfo {author} {\bibfnamefont {J.~M.}\ \bibnamefont {Lattimer}},\
  }\href {\doibase 10.1103/PhysRevD.58.013009} {\bibfield  {journal} {\bibinfo
  {journal} {Phys. Rev. D}\ }\textbf {\bibinfo {volume} {58}},\ \bibinfo
  {pages} {013009} (\bibinfo {year} {1998})}\BibitemShut {NoStop}%
\bibitem [{\citenamefont {Reddy}\ \emph {et~al.}(1999)\citenamefont {Reddy},
  \citenamefont {Prakash}, \citenamefont {Lattimer},\ and\ \citenamefont
  {Pons}}]{Reddy:1999hb}%
  \BibitemOpen
  \bibfield  {author} {\bibinfo {author} {\bibfnamefont {S.}~\bibnamefont
  {Reddy}}, \bibinfo {author} {\bibfnamefont {M.}~\bibnamefont {Prakash}},
  \bibinfo {author} {\bibfnamefont {J.~M.}\ \bibnamefont {Lattimer}}, \ and\
  \bibinfo {author} {\bibfnamefont {J.~A.}\ \bibnamefont {Pons}},\ }\href
  {\doibase 10.1103/PhysRevC.59.2888} {\bibfield  {journal} {\bibinfo
  {journal} {Phys. Rev. C}\ }\textbf {\bibinfo {volume} {59}},\ \bibinfo
  {pages} {2888} (\bibinfo {year} {1999})}\BibitemShut {NoStop}%
\bibitem [{\citenamefont {Horowitz}(2002)}]{Horowitz:2002}%
  \BibitemOpen
  \bibfield  {author} {\bibinfo {author} {\bibfnamefont {C.~J.}\ \bibnamefont
  {Horowitz}},\ }\href {\doibase 10.1103/PhysRevD.65.043001} {\bibfield
  {journal} {\bibinfo  {journal} {Phys. Rev.}\ }\textbf {\bibinfo {volume} {D
  65}},\ \bibinfo {pages} {043001} (\bibinfo {year} {2002})}\BibitemShut
  {NoStop}%
\bibitem [{\citenamefont {Buras}\ \emph {et~al.}(2003)\citenamefont {Buras},
  \citenamefont {Janka}, \citenamefont {Keil}, \citenamefont {Raffelt},\ and\
  \citenamefont {Rampp}}]{Buras:2003wt}%
  \BibitemOpen
  \bibfield  {author} {\bibinfo {author} {\bibfnamefont {R.}~\bibnamefont
  {Buras}}, \bibinfo {author} {\bibfnamefont {H.-T.}\ \bibnamefont {Janka}},
  \bibinfo {author} {\bibfnamefont {M.~T.}\ \bibnamefont {Keil}}, \bibinfo
  {author} {\bibfnamefont {G.~G.}\ \bibnamefont {Raffelt}}, \ and\ \bibinfo
  {author} {\bibfnamefont {M.}~\bibnamefont {Rampp}},\ }\href {\doibase
  10.1086/368015} {\bibfield  {journal} {\bibinfo  {journal} {Astrophys. J.}\
  }\textbf {\bibinfo {volume} {587}},\ \bibinfo {pages} {320} (\bibinfo {year}
  {2003})}\BibitemShut {NoStop}%
\bibitem [{\citenamefont {Horowitz}\ and\ \citenamefont
  {Schwenk}(2006)}]{Horowitz.Schwenk:2006}%
  \BibitemOpen
  \bibfield  {author} {\bibinfo {author} {\bibfnamefont {C.~J.}\ \bibnamefont
  {Horowitz}}\ and\ \bibinfo {author} {\bibfnamefont {A.}~\bibnamefont
  {Schwenk}},\ }\href {\doibase 10.1016/j.physletb.2006.09.042} {\bibfield
  {journal} {\bibinfo  {journal} {Phys. Lett. B}\ }\textbf {\bibinfo {volume}
  {642}},\ \bibinfo {pages} {326} (\bibinfo {year} {2006})}\BibitemShut
  {NoStop}%
\bibitem [{\citenamefont {Juodagalvis}\ \emph {et~al.}(2010)\citenamefont
  {Juodagalvis}, \citenamefont {Langanke}, \citenamefont {Hix}, \citenamefont
  {Mart{\'i}nez-Pinedo},\ and\ \citenamefont {Sampaio}}]{Juodagalvis:2010pt}%
  \BibitemOpen
  \bibfield  {author} {\bibinfo {author} {\bibfnamefont {A.}~\bibnamefont
  {Juodagalvis}}, \bibinfo {author} {\bibfnamefont {K.}~\bibnamefont
  {Langanke}}, \bibinfo {author} {\bibfnamefont {W.~R.}\ \bibnamefont {Hix}},
  \bibinfo {author} {\bibfnamefont {G.}~\bibnamefont {Mart{\'i}nez-Pinedo}}, \
  and\ \bibinfo {author} {\bibfnamefont {J.}~\bibnamefont {Sampaio}},\ }\href
  {\doibase 10.1016/j.nuclphysa.2010.09.012} {\bibfield  {journal} {\bibinfo
  {journal} {Nucl. Phys.}\ }\textbf {\bibinfo {volume} {A848}},\ \bibinfo
  {pages} {454} (\bibinfo {year} {2010})}\BibitemShut {NoStop}%
\bibitem [{\citenamefont {{Roberts}}\ \emph {et~al.}(2012)\citenamefont
  {{Roberts}}, \citenamefont {{Reddy}},\ and\ \citenamefont
  {{Shen}}}]{Roberts.Reddy.Shen:2012}%
  \BibitemOpen
  \bibfield  {author} {\bibinfo {author} {\bibfnamefont {L.~F.}\ \bibnamefont
  {{Roberts}}}, \bibinfo {author} {\bibfnamefont {S.}~\bibnamefont {{Reddy}}},
  \ and\ \bibinfo {author} {\bibfnamefont {G.}~\bibnamefont {{Shen}}},\ }\href
  {\doibase 10.1103/PhysRevC.86.065803} {\bibfield  {journal} {\bibinfo
  {journal} {Phys. Rev. C}\ }\textbf {\bibinfo {volume} {86}},\ \bibinfo {eid}
  {065803} (\bibinfo {year} {2012})}\BibitemShut {NoStop}%
\bibitem [{\citenamefont {Bartl}\ \emph {et~al.}(2014)\citenamefont {Bartl},
  \citenamefont {Pethick},\ and\ \citenamefont {Schwenk}}]{Bartl:2014hoa}%
  \BibitemOpen
  \bibfield  {author} {\bibinfo {author} {\bibfnamefont {A.}~\bibnamefont
  {Bartl}}, \bibinfo {author} {\bibfnamefont {C.~J.}\ \bibnamefont {Pethick}},
  \ and\ \bibinfo {author} {\bibfnamefont {A.}~\bibnamefont {Schwenk}},\ }\href
  {\doibase 10.1103/PhysRevLett.113.081101} {\bibfield  {journal} {\bibinfo
  {journal} {Phys. Rev. Lett.}\ }\textbf {\bibinfo {volume} {113}},\ \bibinfo
  {pages} {081101} (\bibinfo {year} {2014})}\BibitemShut {NoStop}%
\bibitem [{\citenamefont {Horowitz}\ \emph {et~al.}(2017)\citenamefont
  {Horowitz}, \citenamefont {Caballero}, \citenamefont {Lin}, \citenamefont
  {O'Connor},\ and\ \citenamefont {Schwenk}}]{Horowitz.Caballero.ea:2017}%
  \BibitemOpen
  \bibfield  {author} {\bibinfo {author} {\bibfnamefont {C.~J.}\ \bibnamefont
  {Horowitz}}, \bibinfo {author} {\bibfnamefont {O.~L.}\ \bibnamefont
  {Caballero}}, \bibinfo {author} {\bibfnamefont {Z.}~\bibnamefont {Lin}},
  \bibinfo {author} {\bibfnamefont {E.}~\bibnamefont {O'Connor}}, \ and\
  \bibinfo {author} {\bibfnamefont {A.}~\bibnamefont {Schwenk}},\ }\href
  {\doibase 10.1103/PhysRevC.95.025801} {\bibfield  {journal} {\bibinfo
  {journal} {Phys. Rev. C}\ }\textbf {\bibinfo {volume} {95}},\ \bibinfo
  {pages} {025801} (\bibinfo {year} {2017})}\BibitemShut {NoStop}%
\bibitem [{\citenamefont {Roberts}\ and\ \citenamefont
  {Reddy}(2017)}]{Roberts.Reddy:2017}%
  \BibitemOpen
  \bibfield  {author} {\bibinfo {author} {\bibfnamefont {L.~F.}\ \bibnamefont
  {Roberts}}\ and\ \bibinfo {author} {\bibfnamefont {S.}~\bibnamefont
  {Reddy}},\ }\href {\doibase 10.1103/PhysRevC.95.045807} {\bibfield  {journal}
  {\bibinfo  {journal} {Phys. Rev. C}\ }\textbf {\bibinfo {volume} {95}},\
  \bibinfo {pages} {045807} (\bibinfo {year} {2017})}\BibitemShut {NoStop}%
\bibitem [{\citenamefont {Bedaque}\ \emph {et~al.}(2018)\citenamefont
  {Bedaque}, \citenamefont {Reddy}, \citenamefont {Sen},\ and\ \citenamefont
  {Warrington}}]{Bedaque:2018wns}%
  \BibitemOpen
  \bibfield  {author} {\bibinfo {author} {\bibfnamefont {P.~F.}\ \bibnamefont
  {Bedaque}}, \bibinfo {author} {\bibfnamefont {S.}~\bibnamefont {Reddy}},
  \bibinfo {author} {\bibfnamefont {S.}~\bibnamefont {Sen}}, \ and\ \bibinfo
  {author} {\bibfnamefont {N.~C.}\ \bibnamefont {Warrington}},\ }\href
  {\doibase 10.1103/PhysRevC.98.015802} {\bibfield  {journal} {\bibinfo
  {journal} {Phys. Rev. C}\ }\textbf {\bibinfo {volume} {98}},\ \bibinfo
  {pages} {015802} (\bibinfo {year} {2018})}\BibitemShut {NoStop}%
\bibitem [{\citenamefont {Guo}\ and\ \citenamefont
  {Mart{\'i}nez-Pinedo}(2019)}]{Guo:2019cvs}%
  \BibitemOpen
  \bibfield  {author} {\bibinfo {author} {\bibfnamefont {G.}~\bibnamefont
  {Guo}}\ and\ \bibinfo {author} {\bibfnamefont {G.}~\bibnamefont
  {Mart{\'i}nez-Pinedo}},\ }\href {\doibase 10.3847/1538-4357/ab536d}
  {\bibfield  {journal} {\bibinfo  {journal} {Astrophys. J.}\ }\textbf
  {\bibinfo {volume} {887}},\ \bibinfo {pages} {58} (\bibinfo {year}
  {2019})}\BibitemShut {NoStop}%
\bibitem [{\citenamefont {Rampp}\ and\ \citenamefont
  {Janka}(2002)}]{Rampp:2002bq}%
  \BibitemOpen
  \bibfield  {author} {\bibinfo {author} {\bibfnamefont {M.}~\bibnamefont
  {Rampp}}\ and\ \bibinfo {author} {\bibfnamefont {H.-T.}\ \bibnamefont
  {Janka}},\ }\href {\doibase 10.1051/0004-6361:20021398} {\bibfield  {journal}
  {\bibinfo  {journal} {Astron. Astrophys.}\ }\textbf {\bibinfo {volume}
  {396}},\ \bibinfo {pages} {361} (\bibinfo {year} {2002})}\BibitemShut
  {NoStop}%
\bibitem [{\citenamefont {Liebendoerfer}\ \emph {et~al.}(2005)\citenamefont
  {Liebendoerfer}, \citenamefont {Rampp}, \citenamefont {Janka},\ and\
  \citenamefont {Mezzacappa}}]{Liebendoerfer:2005es}%
  \BibitemOpen
  \bibfield  {author} {\bibinfo {author} {\bibfnamefont {M.}~\bibnamefont
  {Liebendoerfer}}, \bibinfo {author} {\bibfnamefont {M.}~\bibnamefont
  {Rampp}}, \bibinfo {author} {\bibfnamefont {H.-T.}\ \bibnamefont {Janka}}, \
  and\ \bibinfo {author} {\bibfnamefont {A.}~\bibnamefont {Mezzacappa}},\
  }\href {\doibase 10.1086/427203} {\bibfield  {journal} {\bibinfo  {journal}
  {Astrophys. J.}\ }\textbf {\bibinfo {volume} {620}},\ \bibinfo {pages} {840}
  (\bibinfo {year} {2005})}\BibitemShut {NoStop}%
\bibitem [{\citenamefont {Buras}\ \emph {et~al.}(2006)\citenamefont {Buras},
  \citenamefont {Rampp}, \citenamefont {Janka},\ and\ \citenamefont
  {Kifonidis}}]{Buras:2006rp}%
  \BibitemOpen
  \bibfield  {author} {\bibinfo {author} {\bibfnamefont {R.}~\bibnamefont
  {Buras}}, \bibinfo {author} {\bibfnamefont {M.}~\bibnamefont {Rampp}},
  \bibinfo {author} {\bibfnamefont {H.-T.}\ \bibnamefont {Janka}}, \ and\
  \bibinfo {author} {\bibfnamefont {K.}~\bibnamefont {Kifonidis}},\ }\href
  {\doibase 10.1051/0004-6361:20053783} {\bibfield  {journal} {\bibinfo
  {journal} {Astron. Astrophys.}\ }\textbf {\bibinfo {volume} {447}},\ \bibinfo
  {pages} {1049} (\bibinfo {year} {2006})}\BibitemShut {NoStop}%
\bibitem [{\citenamefont {Fischer}\ \emph {et~al.}(2010)\citenamefont
  {Fischer}, \citenamefont {Whitehouse}, \citenamefont {Mezzacappa},
  \citenamefont {Thielemann},\ and\ \citenamefont
  {Liebendörfer}}]{Fischer_2010}%
  \BibitemOpen
  \bibfield  {author} {\bibinfo {author} {\bibfnamefont {T.}~\bibnamefont
  {Fischer}}, \bibinfo {author} {\bibfnamefont {S.~C.}\ \bibnamefont
  {Whitehouse}}, \bibinfo {author} {\bibfnamefont {A.}~\bibnamefont
  {Mezzacappa}}, \bibinfo {author} {\bibfnamefont {F.-K.}\ \bibnamefont
  {Thielemann}}, \ and\ \bibinfo {author} {\bibfnamefont {M.}~\bibnamefont
  {Liebendörfer}},\ }\href {\doibase 10.1051/0004-6361/200913106} {\bibfield
  {journal} {\bibinfo  {journal} {Astron. Astrophys.}\ }\textbf {\bibinfo
  {volume} {517}},\ \bibinfo {pages} {A80} (\bibinfo {year}
  {2010})}\BibitemShut {NoStop}%
\bibitem [{\citenamefont {Mart{\'i}nez-Pinedo}\ \emph
  {et~al.}(2012)\citenamefont {Mart{\'i}nez-Pinedo}, \citenamefont {Fischer},
  \citenamefont {Lohs},\ and\ \citenamefont {Huther}}]{MartinezPinedo:2012rb}%
  \BibitemOpen
  \bibfield  {author} {\bibinfo {author} {\bibfnamefont {G.}~\bibnamefont
  {Mart{\'i}nez-Pinedo}}, \bibinfo {author} {\bibfnamefont {T.}~\bibnamefont
  {Fischer}}, \bibinfo {author} {\bibfnamefont {A.}~\bibnamefont {Lohs}}, \
  and\ \bibinfo {author} {\bibfnamefont {L.}~\bibnamefont {Huther}},\ }\href
  {\doibase 10.1103/PhysRevLett.109.251104} {\bibfield  {journal} {\bibinfo
  {journal} {Phys. Rev. Lett.}\ }\textbf {\bibinfo {volume} {109}},\ \bibinfo
  {pages} {251104} (\bibinfo {year} {2012})}\BibitemShut {NoStop}%
\bibitem [{\citenamefont {O'Connor}(2015)}]{OConnor:2015sgn}%
  \BibitemOpen
  \bibfield  {author} {\bibinfo {author} {\bibfnamefont {E.}~\bibnamefont
  {O'Connor}},\ }\href {\doibase 10.1088/0067-0049/219/2/24} {\bibfield
  {journal} {\bibinfo  {journal} {Astrophys. J. Suppl.}\ }\textbf {\bibinfo
  {volume} {219}},\ \bibinfo {pages} {24} (\bibinfo {year} {2015})}\BibitemShut
  {NoStop}%
\bibitem [{\citenamefont {Bartl}\ \emph {et~al.}(2016)\citenamefont {Bartl},
  \citenamefont {Bollig}, \citenamefont {Janka},\ and\ \citenamefont
  {Schwenk}}]{Bartl.Bollig.ea:2016}%
  \BibitemOpen
  \bibfield  {author} {\bibinfo {author} {\bibfnamefont {A.}~\bibnamefont
  {Bartl}}, \bibinfo {author} {\bibfnamefont {R.}~\bibnamefont {Bollig}},
  \bibinfo {author} {\bibfnamefont {H.-T.}\ \bibnamefont {Janka}}, \ and\
  \bibinfo {author} {\bibfnamefont {A.}~\bibnamefont {Schwenk}},\ }\href
  {\doibase 10.1103/PhysRevD.94.083009} {\bibfield  {journal} {\bibinfo
  {journal} {Phys. Rev. D}\ }\textbf {\bibinfo {volume} {94}},\ \bibinfo
  {pages} {083009} (\bibinfo {year} {2016})}\BibitemShut {NoStop}%
\bibitem [{\citenamefont {Fischer}(2016)}]{Fischer:2016boc}%
  \BibitemOpen
  \bibfield  {author} {\bibinfo {author} {\bibfnamefont {T.}~\bibnamefont
  {Fischer}},\ }\href {\doibase 10.1051/0004-6361/201628991} {\bibfield
  {journal} {\bibinfo  {journal} {Astron. Astrophys.}\ }\textbf {\bibinfo
  {volume} {593}},\ \bibinfo {pages} {A103} (\bibinfo {year}
  {2016})}\BibitemShut {NoStop}%
\bibitem [{\citenamefont {Roberts}\ \emph {et~al.}(2016)\citenamefont
  {Roberts}, \citenamefont {Ott}, \citenamefont {Haas}, \citenamefont
  {O'Connor}, \citenamefont {Diener},\ and\ \citenamefont
  {Schnetter}}]{Roberts:2016lzn}%
  \BibitemOpen
  \bibfield  {author} {\bibinfo {author} {\bibfnamefont {L.~F.}\ \bibnamefont
  {Roberts}}, \bibinfo {author} {\bibfnamefont {C.~D.}\ \bibnamefont {Ott}},
  \bibinfo {author} {\bibfnamefont {R.}~\bibnamefont {Haas}}, \bibinfo {author}
  {\bibfnamefont {E.~P.}\ \bibnamefont {O'Connor}}, \bibinfo {author}
  {\bibfnamefont {P.}~\bibnamefont {Diener}}, \ and\ \bibinfo {author}
  {\bibfnamefont {E.}~\bibnamefont {Schnetter}},\ }\href {\doibase
  10.3847/0004-637X/831/1/98} {\bibfield  {journal} {\bibinfo  {journal}
  {Astrophys. J.}\ }\textbf {\bibinfo {volume} {831}},\ \bibinfo {pages} {98}
  (\bibinfo {year} {2016})}\BibitemShut {NoStop}%
\bibitem [{\citenamefont {Kotake}\ \emph {et~al.}(2018)\citenamefont {Kotake},
  \citenamefont {Takiwaki}, \citenamefont {Fischer}, \citenamefont {Nakamura},\
  and\ \citenamefont {Mart{\'i}nez-Pinedo}}]{Kotake:2018ypf}%
  \BibitemOpen
  \bibfield  {author} {\bibinfo {author} {\bibfnamefont {K.}~\bibnamefont
  {Kotake}}, \bibinfo {author} {\bibfnamefont {T.}~\bibnamefont {Takiwaki}},
  \bibinfo {author} {\bibfnamefont {T.}~\bibnamefont {Fischer}}, \bibinfo
  {author} {\bibfnamefont {K.}~\bibnamefont {Nakamura}}, \ and\ \bibinfo
  {author} {\bibfnamefont {G.}~\bibnamefont {Mart{\'i}nez-Pinedo}},\ }\href
  {\doibase 10.3847/1538-4357/aaa716} {\bibfield  {journal} {\bibinfo
  {journal} {Astrophys. J.}\ }\textbf {\bibinfo {volume} {853}},\ \bibinfo
  {pages} {170} (\bibinfo {year} {2018})}\BibitemShut {NoStop}%
\bibitem [{\citenamefont {Fischer}\ \emph {et~al.}(2020)\citenamefont
  {Fischer}, \citenamefont {Guo}, \citenamefont {Dzhioev}, \citenamefont
  {Mart{\'i}nez-Pinedo}, \citenamefont {Wu}, \citenamefont {Lohs},\ and\
  \citenamefont {Qian}}]{Fischer:2020kdt}%
  \BibitemOpen
  \bibfield  {author} {\bibinfo {author} {\bibfnamefont {T.}~\bibnamefont
  {Fischer}}, \bibinfo {author} {\bibfnamefont {G.}~\bibnamefont {Guo}},
  \bibinfo {author} {\bibfnamefont {A.~A.}\ \bibnamefont {Dzhioev}}, \bibinfo
  {author} {\bibfnamefont {G.}~\bibnamefont {Mart{\'i}nez-Pinedo}}, \bibinfo
  {author} {\bibfnamefont {M.-R.}\ \bibnamefont {Wu}}, \bibinfo {author}
  {\bibfnamefont {A.}~\bibnamefont {Lohs}}, \ and\ \bibinfo {author}
  {\bibfnamefont {Y.-Z.}\ \bibnamefont {Qian}},\ }\href {\doibase
  10.1103/PhysRevC.101.025804} {\bibfield  {journal} {\bibinfo  {journal}
  {Phys. Rev. C}\ }\textbf {\bibinfo {volume} {101}},\ \bibinfo {pages}
  {025804} (\bibinfo {year} {2020})}\BibitemShut {NoStop}%
\bibitem [{\citenamefont {Bollig}\ \emph {et~al.}(2017)\citenamefont {Bollig},
  \citenamefont {Janka}, \citenamefont {Lohs}, \citenamefont
  {Mart{\'i}nez-Pinedo}, \citenamefont {Horowitz},\ and\ \citenamefont
  {Melson}}]{Bollig:2017}%
  \BibitemOpen
  \bibfield  {author} {\bibinfo {author} {\bibfnamefont {R.}~\bibnamefont
  {Bollig}}, \bibinfo {author} {\bibfnamefont {H.-T.}\ \bibnamefont {Janka}},
  \bibinfo {author} {\bibfnamefont {A.}~\bibnamefont {Lohs}}, \bibinfo {author}
  {\bibfnamefont {G.}~\bibnamefont {Mart{\'i}nez-Pinedo}}, \bibinfo {author}
  {\bibfnamefont {C.~J.}\ \bibnamefont {Horowitz}}, \ and\ \bibinfo {author}
  {\bibfnamefont {T.}~\bibnamefont {Melson}},\ }\href {\doibase
  10.1103/PhysRevLett.119.242702} {\bibfield  {journal} {\bibinfo  {journal}
  {Phys. Rev. Lett.}\ }\textbf {\bibinfo {volume} {119}},\ \bibinfo {pages}
  {242702} (\bibinfo {year} {2017})}\BibitemShut {NoStop}%
\bibitem [{\citenamefont {Burrows}\ and\ \citenamefont
  {Sawyer}(1998)}]{Burrows:1998cg}%
  \BibitemOpen
  \bibfield  {author} {\bibinfo {author} {\bibfnamefont {A.}~\bibnamefont
  {Burrows}}\ and\ \bibinfo {author} {\bibfnamefont {R.~F.}\ \bibnamefont
  {Sawyer}},\ }\href {\doibase 10.1103/PhysRevC.58.554} {\bibfield  {journal}
  {\bibinfo  {journal} {Phys. Rev. C}\ }\textbf {\bibinfo {volume} {58}},\
  \bibinfo {pages} {554} (\bibinfo {year} {1998})}\BibitemShut {NoStop}%
\bibitem [{\citenamefont {Burrows}\ and\ \citenamefont
  {Sawyer}(1999)}]{Burrows:1999ek}%
  \BibitemOpen
  \bibfield  {author} {\bibinfo {author} {\bibfnamefont {A.}~\bibnamefont
  {Burrows}}\ and\ \bibinfo {author} {\bibfnamefont {R.~F.}\ \bibnamefont
  {Sawyer}},\ }\href {\doibase 10.1103/PhysRevC.59.510} {\bibfield  {journal}
  {\bibinfo  {journal} {Phys. Rev. C}\ }\textbf {\bibinfo {volume} {59}},\
  \bibinfo {pages} {510} (\bibinfo {year} {1999})}\BibitemShut {NoStop}%
\bibitem [{\citenamefont {Horowitz}\ and\ \citenamefont
  {Perez-Garcia}(2003)}]{Horowitz:2003yx}%
  \BibitemOpen
  \bibfield  {author} {\bibinfo {author} {\bibfnamefont {C.~J.}\ \bibnamefont
  {Horowitz}}\ and\ \bibinfo {author} {\bibfnamefont {M.~A.}\ \bibnamefont
  {Perez-Garcia}},\ }\href {\doibase 10.1103/PhysRevC.68.025803} {\bibfield
  {journal} {\bibinfo  {journal} {Phys. Rev. C}\ }\textbf {\bibinfo {volume}
  {68}},\ \bibinfo {pages} {025803} (\bibinfo {year} {2003})}\BibitemShut
  {NoStop}%
\bibitem [{\citenamefont {Fore}\ and\ \citenamefont
  {Reddy}(2020)}]{Fore:2019wib}%
  \BibitemOpen
  \bibfield  {author} {\bibinfo {author} {\bibfnamefont {B.}~\bibnamefont
  {Fore}}\ and\ \bibinfo {author} {\bibfnamefont {S.}~\bibnamefont {Reddy}},\
  }\href {\doibase 10.1103/PhysRevC.101.035809} {\bibfield  {journal} {\bibinfo
   {journal} {Phys. Rev. C}\ }\textbf {\bibinfo {volume} {101}},\ \bibinfo
  {pages} {035809} (\bibinfo {year} {2020})}\BibitemShut {NoStop}%
\bibitem [{\citenamefont {Yueh}\ and\ \citenamefont
  {Buchler}(1976{\natexlab{a}})}]{Yueh:1976A}%
  \BibitemOpen
  \bibfield  {author} {\bibinfo {author} {\bibfnamefont {W.~R.}\ \bibnamefont
  {Yueh}}\ and\ \bibinfo {author} {\bibfnamefont {J.~R.}\ \bibnamefont
  {Buchler}},\ }\href {\doibase 10.1007/BF00648341} {\bibfield  {journal}
  {\bibinfo  {journal} {Astrophys. Space Sci.}\ }\textbf {\bibinfo {volume}
  {39}},\ \bibinfo {pages} {429} (\bibinfo {year}
  {1976}{\natexlab{a}})}\BibitemShut {NoStop}%
\bibitem [{\citenamefont {Yueh}\ and\ \citenamefont
  {Buchler}(1976{\natexlab{b}})}]{Yueh:1976B}%
  \BibitemOpen
  \bibfield  {author} {\bibinfo {author} {\bibfnamefont {W.~R.}\ \bibnamefont
  {Yueh}}\ and\ \bibinfo {author} {\bibfnamefont {J.~R.}\ \bibnamefont
  {Buchler}},\ }\href {\doibase 10.1007/BF00684583} {\bibfield  {journal}
  {\bibinfo  {journal} {Astrophys. Space Sci.}\ }\textbf {\bibinfo {volume}
  {41}},\ \bibinfo {pages} {221} (\bibinfo {year}
  {1976}{\natexlab{b}})}\BibitemShut {NoStop}%
\bibitem [{\citenamefont {Lohs}()}]{Lohs:2015}%
  \BibitemOpen
  \bibfield  {author} {\bibinfo {author} {\bibfnamefont {A.}~\bibnamefont
  {Lohs}},\ }\href@noop {} {Ph.D. thesis},\ \bibinfo  {school} {Technische
  Universit{\"a}t Darmstadt, 2015}\BibitemShut {NoStop}%
\bibitem [{\citenamefont {Fuller}\ \emph {et~al.}(1985)\citenamefont {Fuller},
  \citenamefont {Fowler},\ and\ \citenamefont {Newman}}]{Fuller:1085zz}%
  \BibitemOpen
  \bibfield  {author} {\bibinfo {author} {\bibfnamefont {G.}~\bibnamefont
  {Fuller}}, \bibinfo {author} {\bibfnamefont {W.}~\bibnamefont {Fowler}}, \
  and\ \bibinfo {author} {\bibfnamefont {M.~J.}\ \bibnamefont {Newman}},\
  }\href {\doibase 10.1086/163208} {\bibfield  {journal} {\bibinfo  {journal}
  {Astrophys. J.}\ }\textbf {\bibinfo {volume} {293}},\ \bibinfo {pages} {1}
  (\bibinfo {year} {1985})}\BibitemShut {NoStop}%
\bibitem [{\citenamefont {Mart{\'i}nez-Pinedo}\ \emph
  {et~al.}(2014)\citenamefont {Mart{\'i}nez-Pinedo}, \citenamefont {Lam},
  \citenamefont {Langanke}, \citenamefont {Zegers},\ and\ \citenamefont
  {Sullivan}}]{Martinez-Pinedo:2014koa}%
  \BibitemOpen
  \bibfield  {author} {\bibinfo {author} {\bibfnamefont {G.}~\bibnamefont
  {Mart{\'i}nez-Pinedo}}, \bibinfo {author} {\bibfnamefont {Y.~H.}\
  \bibnamefont {Lam}}, \bibinfo {author} {\bibfnamefont {K.}~\bibnamefont
  {Langanke}}, \bibinfo {author} {\bibfnamefont {R.~G.~T.}\ \bibnamefont
  {Zegers}}, \ and\ \bibinfo {author} {\bibfnamefont {C.}~\bibnamefont
  {Sullivan}},\ }\href {\doibase 10.1103/PhysRevC.89.045806} {\bibfield
  {journal} {\bibinfo  {journal} {Phys. Rev. C}\ }\textbf {\bibinfo {volume}
  {89}},\ \bibinfo {pages} {045806} (\bibinfo {year} {2014})}\BibitemShut
  {NoStop}%
\bibitem [{\citenamefont {Shen}\ \emph {et~al.}(1998)\citenamefont {Shen},
  \citenamefont {Toki}, \citenamefont {Oyamatsu},\ and\ \citenamefont
  {Sumiyoshi}}]{Shen:1998gq}%
  \BibitemOpen
  \bibfield  {author} {\bibinfo {author} {\bibfnamefont {H.}~\bibnamefont
  {Shen}}, \bibinfo {author} {\bibfnamefont {H.}~\bibnamefont {Toki}}, \bibinfo
  {author} {\bibfnamefont {K.}~\bibnamefont {Oyamatsu}}, \ and\ \bibinfo
  {author} {\bibfnamefont {K.}~\bibnamefont {Sumiyoshi}},\ }\href {\doibase
  10.1016/S0375-9474(98)00236-X} {\bibfield  {journal} {\bibinfo  {journal}
  {Nucl. Phys.}\ }\textbf {\bibinfo {volume} {A637}},\ \bibinfo {pages} {435}
  (\bibinfo {year} {1998})}\BibitemShut {NoStop}%
\bibitem [{\citenamefont {Typel}\ and\ \citenamefont
  {Wolter}(1999)}]{Typel:1999yq}%
  \BibitemOpen
  \bibfield  {author} {\bibinfo {author} {\bibfnamefont {S.}~\bibnamefont
  {Typel}}\ and\ \bibinfo {author} {\bibfnamefont {H.~H.}\ \bibnamefont
  {Wolter}},\ }\href {\doibase 10.1016/S0375-9474(99)00310-3} {\bibfield
  {journal} {\bibinfo  {journal} {Nucl. Phys.}\ }\textbf {\bibinfo {volume}
  {A656}},\ \bibinfo {pages} {331} (\bibinfo {year} {1999})}\BibitemShut
  {NoStop}%
\bibitem [{\citenamefont {Typel}(2005)}]{Typel:2005ba}%
  \BibitemOpen
  \bibfield  {author} {\bibinfo {author} {\bibfnamefont {S.}~\bibnamefont
  {Typel}},\ }\href {\doibase 10.1103/PhysRevC.71.064301} {\bibfield  {journal}
  {\bibinfo  {journal} {Phys. Rev. C}\ }\textbf {\bibinfo {volume} {71}},\
  \bibinfo {pages} {064301} (\bibinfo {year} {2005})}\BibitemShut {NoStop}%
\bibitem [{\citenamefont {Hempel}\ and\ \citenamefont
  {Schaffner-Bielich}(2010)}]{Hempel:2010mc}%
  \BibitemOpen
  \bibfield  {author} {\bibinfo {author} {\bibfnamefont {M.}~\bibnamefont
  {Hempel}}\ and\ \bibinfo {author} {\bibfnamefont {J.}~\bibnamefont
  {Schaffner-Bielich}},\ }\href {\doibase 10.1016/j.nuclphysa.2010.02.010}
  {\bibfield  {journal} {\bibinfo  {journal} {Nucl. Phys.}\ }\textbf {\bibinfo
  {volume} {A837}},\ \bibinfo {pages} {210} (\bibinfo {year}
  {2010})}\BibitemShut {NoStop}%
\bibitem [{\citenamefont {Shen}\ \emph {et~al.}(2010)\citenamefont {Shen},
  \citenamefont {Horowitz},\ and\ \citenamefont {Teige}}]{Shen:2010pu}%
  \BibitemOpen
  \bibfield  {author} {\bibinfo {author} {\bibfnamefont {G.}~\bibnamefont
  {Shen}}, \bibinfo {author} {\bibfnamefont {C.~J.}\ \bibnamefont {Horowitz}},
  \ and\ \bibinfo {author} {\bibfnamefont {S.}~\bibnamefont {Teige}},\ }\href
  {\doibase 10.1103/PhysRevC.82.015806} {\bibfield  {journal} {\bibinfo
  {journal} {Phys. Rev. C}\ }\textbf {\bibinfo {volume} {82}},\ \bibinfo
  {pages} {015806} (\bibinfo {year} {2010})}\BibitemShut {NoStop}%
\bibitem [{\citenamefont {Typel}\ \emph {et~al.}(2010)\citenamefont {Typel},
  \citenamefont {Ropke}, \citenamefont {Klahn}, \citenamefont {Blaschke},\ and\
  \citenamefont {Wolter}}]{Typel:2010sy}%
  \BibitemOpen
  \bibfield  {author} {\bibinfo {author} {\bibfnamefont {S.}~\bibnamefont
  {Typel}}, \bibinfo {author} {\bibfnamefont {G.}~\bibnamefont {Ropke}},
  \bibinfo {author} {\bibfnamefont {T.}~\bibnamefont {Klahn}}, \bibinfo
  {author} {\bibfnamefont {D.}~\bibnamefont {Blaschke}}, \ and\ \bibinfo
  {author} {\bibfnamefont {H.~H.}\ \bibnamefont {Wolter}},\ }\href {\doibase
  10.1103/PhysRevC.81.015803} {\bibfield  {journal} {\bibinfo  {journal} {Phys.
  Rev. C}\ }\textbf {\bibinfo {volume} {81}},\ \bibinfo {pages} {015803}
  (\bibinfo {year} {2010})}\BibitemShut {NoStop}%
\bibitem [{\citenamefont {Furusawa}\ \emph {et~al.}(2011)\citenamefont
  {Furusawa}, \citenamefont {Yamada}, \citenamefont {Sumiyoshi},\ and\
  \citenamefont {Suzuki}}]{Furusawa:2011wh}%
  \BibitemOpen
  \bibfield  {author} {\bibinfo {author} {\bibfnamefont {S.}~\bibnamefont
  {Furusawa}}, \bibinfo {author} {\bibfnamefont {S.}~\bibnamefont {Yamada}},
  \bibinfo {author} {\bibfnamefont {K.}~\bibnamefont {Sumiyoshi}}, \ and\
  \bibinfo {author} {\bibfnamefont {H.}~\bibnamefont {Suzuki}},\ }\href
  {\doibase 10.1088/0004-637X/738/2/178} {\bibfield  {journal} {\bibinfo
  {journal} {Astrophys. J.}\ }\textbf {\bibinfo {volume} {738}},\ \bibinfo
  {pages} {178} (\bibinfo {year} {2011})}\BibitemShut {NoStop}%
\bibitem [{\citenamefont {Shen}\ \emph
  {et~al.}(2011{\natexlab{a}})\citenamefont {Shen}, \citenamefont {Horowitz},\
  and\ \citenamefont {Teige}}]{Shen:2011kr}%
  \BibitemOpen
  \bibfield  {author} {\bibinfo {author} {\bibfnamefont {G.}~\bibnamefont
  {Shen}}, \bibinfo {author} {\bibfnamefont {C.~J.}\ \bibnamefont {Horowitz}},
  \ and\ \bibinfo {author} {\bibfnamefont {S.}~\bibnamefont {Teige}},\ }\href
  {\doibase 10.1103/PhysRevC.83.035802} {\bibfield  {journal} {\bibinfo
  {journal} {Phys. Rev. C}\ }\textbf {\bibinfo {volume} {83}},\ \bibinfo
  {pages} {035802} (\bibinfo {year} {2011}{\natexlab{a}})}\BibitemShut
  {NoStop}%
\bibitem [{\citenamefont {Shen}\ \emph
  {et~al.}(2011{\natexlab{b}})\citenamefont {Shen}, \citenamefont {Horowitz},\
  and\ \citenamefont {O'Connor}}]{Shen:2011fc}%
  \BibitemOpen
  \bibfield  {author} {\bibinfo {author} {\bibfnamefont {G.}~\bibnamefont
  {Shen}}, \bibinfo {author} {\bibfnamefont {C.~J.}\ \bibnamefont {Horowitz}},
  \ and\ \bibinfo {author} {\bibfnamefont {E.}~\bibnamefont {O'Connor}},\
  }\href {\doibase 10.1103/PhysRevC.83.065808} {\bibfield  {journal} {\bibinfo
  {journal} {Phys. Rev. C}\ }\textbf {\bibinfo {volume} {83}},\ \bibinfo
  {pages} {065808} (\bibinfo {year} {2011}{\natexlab{b}})}\BibitemShut
  {NoStop}%
\bibitem [{\citenamefont {Hempel}\ \emph {et~al.}(2012)\citenamefont {Hempel},
  \citenamefont {Fischer}, \citenamefont {Schaffner-Bielich},\ and\
  \citenamefont {Liebendorfer}}]{Hempel:2012mk}%
  \BibitemOpen
  \bibfield  {author} {\bibinfo {author} {\bibfnamefont {M.}~\bibnamefont
  {Hempel}}, \bibinfo {author} {\bibfnamefont {T.}~\bibnamefont {Fischer}},
  \bibinfo {author} {\bibfnamefont {J.}~\bibnamefont {Schaffner-Bielich}}, \
  and\ \bibinfo {author} {\bibfnamefont {M.}~\bibnamefont {Liebendorfer}},\
  }\href {\doibase 10.1088/0004-637X/748/1/70} {\bibfield  {journal} {\bibinfo
  {journal} {Astrophys. J.}\ }\textbf {\bibinfo {volume} {748}},\ \bibinfo
  {pages} {70} (\bibinfo {year} {2012})}\BibitemShut {NoStop}%
\bibitem [{\citenamefont {Steiner}\ \emph {et~al.}(2013)\citenamefont
  {Steiner}, \citenamefont {Hempel},\ and\ \citenamefont
  {Fischer}}]{Steiner:2013rk}%
  \BibitemOpen
  \bibfield  {author} {\bibinfo {author} {\bibfnamefont {A.~W.}\ \bibnamefont
  {Steiner}}, \bibinfo {author} {\bibfnamefont {M.}~\bibnamefont {Hempel}}, \
  and\ \bibinfo {author} {\bibfnamefont {T.}~\bibnamefont {Fischer}},\ }\href
  {\doibase 10.1088/0004-637X/774/1/17} {\bibfield  {journal} {\bibinfo
  {journal} {Astrophys. J.}\ }\textbf {\bibinfo {volume} {774}},\ \bibinfo
  {pages} {17} (\bibinfo {year} {2013})}\BibitemShut {NoStop}%
\bibitem [{\citenamefont {Lattimer}\ and\ \citenamefont
  {Swesty}(1991)}]{Lattimer.Swesty:1991}%
  \BibitemOpen
  \bibfield  {author} {\bibinfo {author} {\bibfnamefont {J.~M.}\ \bibnamefont
  {Lattimer}}\ and\ \bibinfo {author} {\bibfnamefont {F.~D.}\ \bibnamefont
  {Swesty}},\ }\href {\doibase https://doi.org/10.1016/0375-9474(91)90452-C}
  {\bibfield  {journal} {\bibinfo  {journal} {Nuclear Physics A}\ }\textbf
  {\bibinfo {volume} {535}},\ \bibinfo {pages} {331 } (\bibinfo {year}
  {1991})}\BibitemShut {NoStop}%
\bibitem [{\citenamefont {Leinson}\ and\ \citenamefont
  {P{\'e}rez}(2001)}]{Leinson_2001}%
  \BibitemOpen
  \bibfield  {author} {\bibinfo {author} {\bibfnamefont {L.~B.}\ \bibnamefont
  {Leinson}}\ and\ \bibinfo {author} {\bibfnamefont {A.}~\bibnamefont
  {P{\'e}rez}},\ }\href {\doibase 10.1016/s0370-2693(01)01042-5} {\bibfield
  {journal} {\bibinfo  {journal} {Physics Letters B}\ }\textbf {\bibinfo
  {volume} {518}},\ \bibinfo {pages} {15} (\bibinfo {year} {2001})}\BibitemShut
  {NoStop}%
\bibitem [{\citenamefont {Leinson}(2002)}]{Leinson:2002bw}%
  \BibitemOpen
  \bibfield  {author} {\bibinfo {author} {\bibfnamefont {L.~B.}\ \bibnamefont
  {Leinson}},\ }\href {\doibase 10.1016/S0375-9474(02)00991-0} {\bibfield
  {journal} {\bibinfo  {journal} {Nucl. Phys.}\ }\textbf {\bibinfo {volume}
  {A707}},\ \bibinfo {pages} {543} (\bibinfo {year} {2002})}\BibitemShut
  {NoStop}%
\bibitem [{\citenamefont {Hahn}(2005)}]{Hahn:2005}%
  \BibitemOpen
  \bibfield  {author} {\bibinfo {author} {\bibfnamefont {T.}~\bibnamefont
  {Hahn}},\ }\href {\doibase 10.1016/j.cpc.2005.01.010} {\bibfield  {journal}
  {\bibinfo  {journal} {Comp. Phys. Comm.}\ }\textbf {\bibinfo {volume}
  {168}},\ \bibinfo {pages} {78} (\bibinfo {year} {2005})}\BibitemShut
  {NoStop}%
\bibitem [{gar()}]{garching}%
  \BibitemOpen
  \href@noop {} {}\bibinfo {howpublished}
  {\url{https://wwwmpa.mpa-garching.mpg.de/ccsnarchive}}\BibitemShut {NoStop}%
\bibitem [{\citenamefont {Woosley}\ and\ \citenamefont
  {Heger}(2007)}]{Woosley.Heger:2007}%
  \BibitemOpen
  \bibfield  {author} {\bibinfo {author} {\bibfnamefont {S.}~\bibnamefont
  {Woosley}}\ and\ \bibinfo {author} {\bibfnamefont {A.}~\bibnamefont
  {Heger}},\ }\href {\doibase 10.1016/j.physrep.2007.02.009} {\bibfield
  {journal} {\bibinfo  {journal} {Phys. Rept.}\ }\textbf {\bibinfo {volume}
  {442}},\ \bibinfo {pages} {269} (\bibinfo {year} {2007})}\BibitemShut
  {NoStop}%
\bibitem [{\citenamefont {{Keil}}\ \emph {et~al.}(2003)\citenamefont {{Keil}},
  \citenamefont {{Raffelt}},\ and\ \citenamefont
  {{Janka}}}]{Keil.Raffelt.Janka:2003}%
  \BibitemOpen
  \bibfield  {author} {\bibinfo {author} {\bibfnamefont {M.~T.}\ \bibnamefont
  {{Keil}}}, \bibinfo {author} {\bibfnamefont {G.~G.}\ \bibnamefont
  {{Raffelt}}}, \ and\ \bibinfo {author} {\bibfnamefont {H.-T.}\ \bibnamefont
  {{Janka}}},\ }\href {\doibase 10.1086/375130} {\bibfield  {journal} {\bibinfo
   {journal} {Astrophys. J.}\ }\textbf {\bibinfo {volume} {590}},\ \bibinfo
  {pages} {971} (\bibinfo {year} {2003})}\BibitemShut {NoStop}%
\end{thebibliography}
\end{document}